\documentclass[twocolumn,trackchanges,floatfix]{aastex7}
\usepackage{booktabs}
\usepackage[english]{babel}
\usepackage{url}
\usepackage{natbib}
\usepackage{gensymb}
\usepackage{subfigure}
\DeclareUnicodeCharacter{2212}{-}
\usepackage{amsmath,bm}
\usepackage{longtable}
\usepackage{multirow}
\usepackage{threeparttable}
\usepackage{times}
\usepackage{enumitem}
\usepackage{lipsum}[1-2]
\usepackage{array}
\usepackage{tabularx}
\usepackage{float}

\shorttitle{Estimating photometric Fe and Mg of dwarf galaxies}
\shortauthors{Hong et al.}
\begin{document}

\title{Estimating Fe and Mg Abundances in the Milky Way Dwarf Galaxies Using Subaru/HSC and DEIMOS}

\author[0000-0002-2453-0853]{Jihye Hong}
\affiliation{Department of Physics and Astronomy, University of Notre Dame, Notre Dame, IN 46556, USA}
\affiliation{Joint Institute for Nuclear Astrophysics -- Center for the Evolution of the Elements (JINA-CEE), USA}
\email{jhong5@nd.edu}

\author[0000-0001-6196-5162]{Evan N. Kirby}
\affiliation{Department of Physics and Astronomy, University of Notre Dame, Notre Dame, IN 46556, USA}
\email{ekirby@nd.edu}

\author[0000-0002-8079-6867]{Tiffany M. Tang}
\affiliation{Department of Applied and Computational Mathematics and Statistics, University of Notre Dame, Notre Dame, IN 46556, USA}
\email{ttang4@nd.edu}

\author[0000-0002-9053-860X]{Masashi Chiba}
\affiliation{Astronomical Institute, Tohoku University, 6-3 Aoba, Aramaki, Aoba-ku, Sendai, Miyagi 980-8578, Japan}
\email{chiba@astr.tohoku.ac.jp}

\author[0000-0002-3852-6329]{Yutaka Komiyama}
\affiliation{Department of Advanced Sciences, Faculty of Science and Engineering, Hosei University, 3-7-2 Kajino-cho, Koganei, Tokyo 184-8584, Japan}
\email{komiyama@hosei.ac.jp}

\author[0000-0001-6207-4388]{Lauren E. Henderson}
\affiliation{Department of Physics and Astronomy, University of Notre Dame, Notre Dame, IN 46556, USA}
\email{lhender6@nd.edu}

\author[0000-0001-8239-4549]{Itsuki Ogami}
\affiliation{National Astronomical Observatory of Japan, 2-21-1 Osawa, Mitaka, Tokyo 181-8588, Japan}
\affiliation{The Institute of Statistical Mathematics, 10-3 Midoricho, Tachikawa, Tokyo, 190-8562, Japan}
\email{itsuki.ogami@nao.ac.jp}

\author[0000-0003-4573-6233]{Timothy C. Beers}
\affiliation{Department of Physics and Astronomy, University of Notre Dame, Notre Dame, IN 46556, USA}
\affiliation{Joint Institute for Nuclear Astrophysics -- Center for the Evolution of the Elements (JINA-CEE), USA}
\email{Timothy.C.Beers.5@nd.edu}

\accepted{Jun17}

\begin{abstract}

We investigate the chemical abundance distributions of the Fornax, Sculptor, Ursa Minor, and Draco dwarf galaxies using Subaru/HSC photometric data. The HSC dataset, which includes broadband $g$ and $i$ filters and the narrowband $NB515$ filter, offers sensitivity to iron and magnesium abundances as well as surface gravity, enabling the identification of giant stars and foreground dwarfs. For analysis, we selected a total of 6713 giant candidates using a Random Forest regressor trained on medium-resolution (R $\sim$ 6000) Keck/DEIMOS spectroscopic data. Our analysis reveals the extent of radial metallicity gradients in the galaxies. Such trends, not detectable in earlier studies, are now captured owing to the substantially enlarged sample size and areal coverage provided by the HSC data. These results are also consistent with chemical abundance patterns previously observed in the central regions through spectroscopic studies. Furthermore, we infer that Fornax underwent extended star formation, whereas Sculptor formed both metal-poor and metal-rich stars over a shorter time. Ursa Minor and Draco appear to have experienced brief, intense star formation episodes leading to nearly extinguished star formation. This study underscores the critical role of the expanded HSC dataset in revealing chemical gradients that were previously inaccessible. Future work incorporating additional spectra of metal-poor stars and age-sensitive isochrone modeling will enable more accurate maps of chemical abundance distributions.

\end{abstract}


\section{Introduction}\label{sec:introduction}

Nearly all of the dwarf galaxies in the vicinity of the Milky Way began forming stars more than 10 billion years ago \citep{Tolstoy2009}. Notable examples include Fornax (Fnx) and Sculptor (Scl) in the southern sky, located about 147 and 86 kpc from the Sun, and Ursa Minor (UMi) and Draco (Dra) in the northern sky, about 76 kpc away (\citealp[and references therein]{McConnachie2012}). Spectroscopic observations and chemical evolution modeling have been extensively applied to explore their star formation histories (SFHs) and the evolution of their chemical properties over time \citep{Kirby2010, Kirby2011b, Kirby2011a, Kirby2013, Ural2015, Reichert2020, Hasselquist2021, Tang2023, Lucchesi2024, Skuladottir2024}.

Medium- and high-resolution spectroscopic surveys provide accurate measurements of elemental abundances of stars in dwarf galaxies. For example, \citet{Kirby2011b,Kirby2011a} analyzed the abundances of Fe, Mg, Si, Ca, and Ti of stars in Fnx. They observed a narrow peak at ${\rm [Fe/H]} = -1.0$ and, using analytic chemical evolution models, predicted the SFH of Fnx: early generations of massive stars enriched the initially pristine gas with metals. Subsequently, stars formed from this enriched interstellar medium, resulting in a higher fraction of metal-rich stars compared to other dwarf galaxies, with these stars more concentrated toward the center. During the SF period, continuous infall of metal-free gas was also required. Additionally, a spike observed in the [$\alpha$/Fe] ratio suggested that SF in Fnx occurred in a bursty and non-uniform manner.

However, the faint light from distant dwarf galaxies limits the number of stars that can be observed spectroscopically. For instance, spectroscopic surveys with a resolving power of around $R \sim 7000$ can observe stars down to a magnitude of $V\sim21.5$ \citep{Kirby2010}. On the other hand, recent photometric surveys cover broader sky areas and allow the detection of fainter stars, significantly expanding the sample size. For instance, the Hyper Suprime-Cam \citep[HSC,][]{Miyazaki2012, Miyazaki2018, Komiyama2018b, Kawanomoto2018, Furusawa2018} mounted on the Subaru Telescope can observe stars down to $r\sim27$~mag\footnote{https://subarutelescope.org//Observing/Instruments/HSC/sensitivity.html}, increasing the sample size by several hundred times. Leveraging stars with spectroscopically measured elemental abundances to estimate the photometric abundances of hundreds of thousands of stars will provide a better understanding of the overall chemical abundance distributions in dwarf galaxies.

The broad $g$- and $i$-band filters of HSC enable us to estimate the [Fe/H] values of stars \citep[e.g.,][]{Okamoto2024}. Additionally, HSC has narrow-band filters, such as $NB515$, which includes the prominent Mg~b triplet of absorption lines at $\lambda \sim 5150$ \AA\@. The strength of Mg~b can be used to estimate [Mg/Fe], which provides clues about the SFH of dwarf galaxies. For example, if metal-poor stars in a dwarf galaxy exhibit high [Mg/Fe] values, it suggests that the galaxy experienced significant core-collapse supernovae (CCSNe) activity early on, driven by massive stars with initial masses of $8-25 M_{\odot}$ because magnesium---an $\alpha$ element---is a typical product of CCSNe. While iron is also produced in CCSNe, it is also generated in Type Ia supernovae (SNe). These events occur hundreds to thousands of Myr after CCSNe \citep{Maoz2010a}, leading to the formation of the so-called $\alpha-$knee, marking the point at which Type Ia SNe become the dominant source of iron enrichment in galaxies \citep{McWilliam1997}. Moreover, since red giant stars in dwarf galaxies tend to exhibit weaker Mg~b triplet absorption lines compared to foreground Milky Way dwarf stars due to their relatively lower surface gravity \citep{Ohman1934, Thackeray1939, Majewski2000}, $NB515$ can distinguish red giant candidates from foreground dwarf stars during the membership assignment process \citep{Komiyama2018a, Ogami2024, Ogami2025} of dwarf galaxy stars.

We estimate the photometric Fe and Mg abundances of 6713 stars by leveraging the advantages of both broad and narrow HSC bands, using a Random Forest (RF) regressor trained on 610 medium-resolution DEep Imaging Multi-Object Spectrograph (DEIMOS) spectroscopic abundance measurements. The details of the data and methods we used are described in Section~\ref{sec:dataandmethods}. In Section~\ref{sec:results}, we present our results, with the analysis and discussion of these findings provided in Section~\ref{sec:discussion}. We conclude with a summary in Section~\ref{sec:summary}.

\section{Data and Methods}\label{sec:dataandmethods}
Here, we detail the data and methods used to estimate the photometric Fe and Mg abundances of individual stars in dwarf galaxies.

\subsection{Data}\label{sec:data}

We used photometry obtained from the HSC mounted on the 8.2 m Subaru Telescope. The observations were conducted in September 2016 for Fnx, in October 2015 and September 2016 for Scl, in May 2015 for UMi, and in April 2016 for Dra. The imaging data of HSC were reduced following the procedure described in \citet{Komiyama2018a}. HSC has a field of view of 1.5 degrees in diameter, allowing it to observe 1.8 deg$^2$ of the sky in a single observation ($g \sim 27.6$~mag). The observations utilized the $g$ and $i$ broadband filters, along with the narrow-band $NB515$ filter, which specifically targets the \ion{Mg}{1}~b triplet and is sensitive to surface gravity \citep{Ohman1934, Thackeray1939, Majewski2000}. Since the giants in dwarf galaxies have lower surface gravity compared to dwarfs in the Milky Way, they exhibit lower (i.e., brighter) magnitude values in the $NB515$ filter, allowing us to distinguish giant members of the dwarf galaxies from foreground dwarfs. We selected Fnx, Scl, UMi, and Dra because they are the classical dSphs with well-calibrated HSC photometry.

\begin{figure*}
    \centering
    \includegraphics[width=\textwidth]{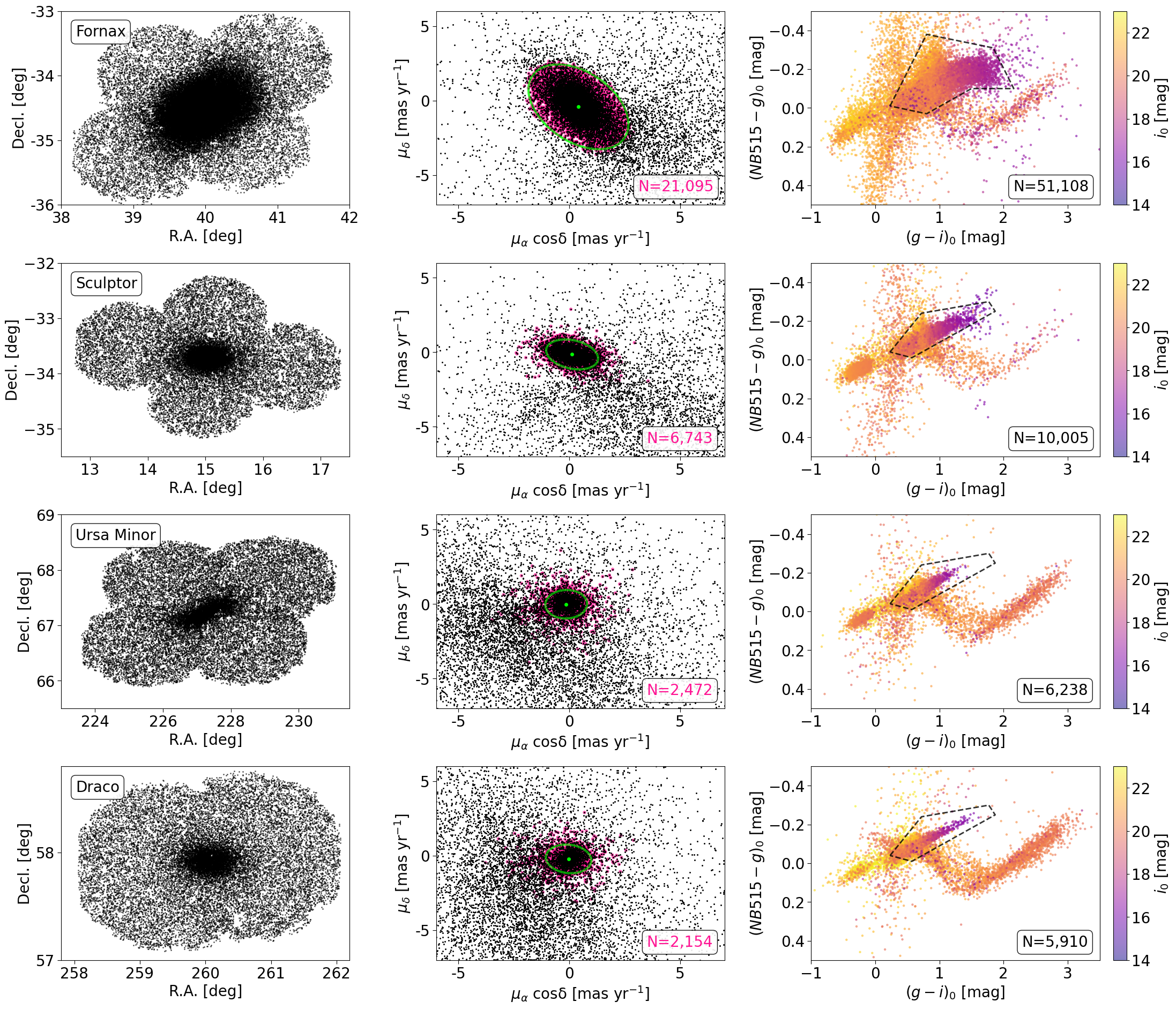}
    \caption{Left: The equatorial coordinates in R.A. and Decl of the dwarf galaxies from the HSC catalog are represented by black points, corresponding to Fnx, Scl, UMi, and Dra from top to bottom, respectively. Middle: Proper motions in in $\mu_{\alpha}$cos$\delta$ and $\mu_{\delta}$ for stars with Gaia DR3 astrometry. The green dot and the ellipse represent the mean proper motion and three times the standard deviation calculated via sigma clipping. Stars marked in pink are either within the ellipse or outside but consistent with the ellipse within their proper motion errors. The number of these stars is indicated in the lower right corner of each panel. Right: A color-color diagram, including the pink stars from the middle panel and stars without Gaia DR3 information. Points are color-coded by their $i_0$ magnitudes. The dashed polygon delineates the giant candidates selected for prediction sample, with their number indicated in the lower right corner.} 
    \label{Fig:ccd_select_sample}
\end{figure*}

We selected point-like sources classified as \texttt{extendedness = 0} in the HSC catalog. Additionally, we applied a magnitude error cut of $\leq 0.02$, which applies to each of $g$, $NB515$, and $i$. The equatorial coordinates of these stars are presented in the left panel of Figure~\ref{Fig:ccd_select_sample}. The proper motions of stars with Gaia DR3 information available are shown as black dots in the middle panel. Stars marked as pink points satisfy the following condition:
\begin{equation} \label{eq:pm}
  \begin{aligned}
r_{b}^2~({\rm pmRA} - {\rm pmRA}_{0})^2 + r_{a}^2~({\rm pmDEC} - {\rm pmDEC}_{0})^2 \\
\leq (r_{a}r_{b})^2 + ({\rm err}_{\rm pm})^2
  \end{aligned}
\end{equation}
where pmRA and pmDEC are the proper motion values of the stars. The pmRA$_{0}$ and pmDEC$_{0}$, marked by the green dot, are the mean values obtained using sigma clipping of \texttt{astropy}, while the $r_{a}$ and $r_{b}$ of green ellipses represent three times the standard deviation derived from the same method ($\sigma = 3$ for Fnx, 2 for Scl, and 1.5 for UMi and Dra). The err$_{\rm pm}$ represents the quadrature sum of the errors in pmRA and pmDEC\@. The right panel shows the color-color diagram of these pink stars, along with stars lacking Gaia astrometry, with points color-coded by the $i$-band magnitudes. From this diagram, we selected giant candidates within the dashed polygon, which was defined with reference to the methods used by \citet{Komiyama2018a} and \citet{Okamoto2024,Ogami2025} for selecting red giant branch (RGB) stars.
The variation in polygon shapes between Fornax and the other three galaxies was designed to exclude potential foreground Milky Way stars associated with a ``$\sim$''-shaped feature. In addition, some main-sequence turn-off (MSTO) stars falling within the range $0.5 < (g-i)_0 < 1.0$ and $-0.1 < (NB515-g)_0 < 0.0$ were enclosed by the selection polygon, but were subsequently excluded through an additional magnitude cut (see below).

The spectra were obtained with the DEIMOS instrument on the Keck II 10 m Telescope. Observations were conducted by \citet{Kirby2010} and by \citet{Henderson2025}. Spectroscopic [Fe/H] and [Mg/H] abundances were measured from these medium-resolution spectra ($R \sim 6000$). The spectroscopic datasets were cross-matched with the HSC catalog using a 1'' radius after removing duplicates, in order to construct the training and test samples. We refer to this cross-matched sample as the DEIMOS-HSC sample hereafter. After applying a magnitude error cut of $\leq 0.02$, we calculated the absolute magnitudes for these stars in both the HSC and DEIMOS datasets using the distance moduli from \citet{McConnachie2012}. The proper motion constraints were applied to the stars with Gaia DR3 information, all of which satisfied Equation~\ref{eq:pm}. Finally, spectroscopically confirmed giant stars were selected from the dashed polygon in the color-color diagram.

\begin{figure*}
    \centering
    \includegraphics[width=0.76\textwidth]{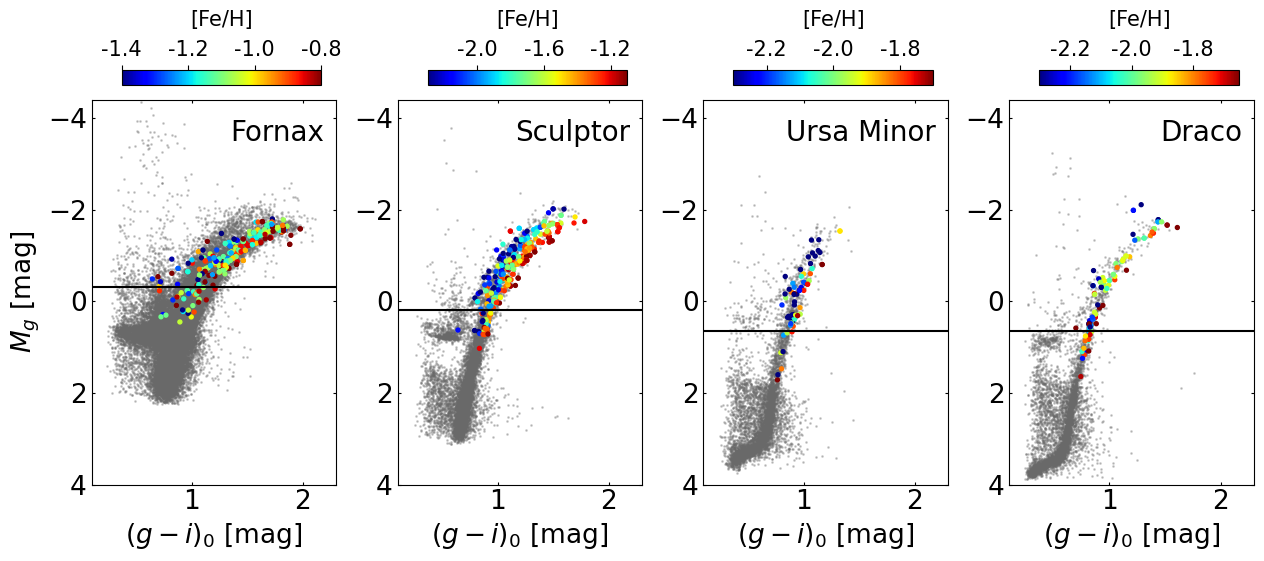}
    \caption{Color-magnitude diagrams for Fnx, Scl, UMi, and Dra, shown from left to right. The gray dots represent the HSC photometry data, while the color-coded dots based on [Fe/H] represent the DEIMOS-HSC sample. The black solid lines indicate the faintest absolute magnitudes that define the final prediction and training samples.}
    \label{Fig:tailoredmag}
\end{figure*}

To enhance the reliability of the spectroscopic abundance measurements, we applied an error cut of [Mg/Fe]$_{\rm err} < 0.3$. To further improve the quality of the DEIMOS-HSC sample, additional cuts were applied to each magnitude, including $M_g$, as shown in Figure~\ref{Fig:tailoredmag}, with the black solid line representing the cut. The points are color-coded by [Fe/H], with the gray points indicating the HSC sample, which were likewise filtered to align with the cuts. At this stage, possible MSTO stars were excluded, leaving giant stars. This process resulted in a total of 610 stars in the spectroscopic DEIMOS-HSC sample (Fnx: 228, Scl: 247, UMi: 58, Dra: 77), of which 80\% was used for training and 20\% used for test, and 9371 stars in the photometric HSC prediction sample (Fnx: 6870, Scl: 1397, UMi: 615, Dra: 489).

\begin{figure}
    \centering
    \includegraphics[width=0.35\textwidth]{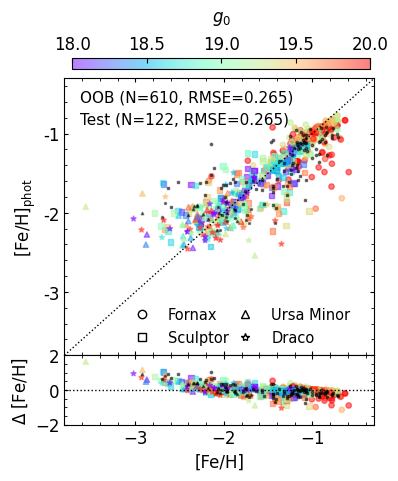}
    \includegraphics[width=0.35\textwidth]{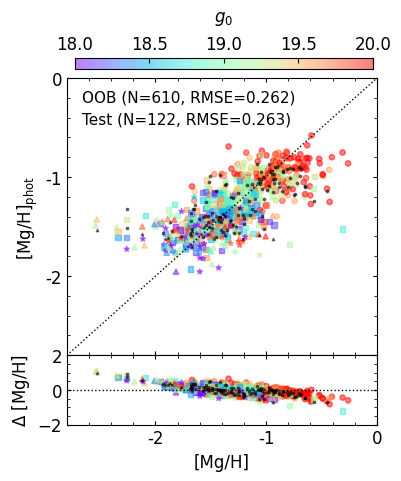}
    \caption{Comparison of measured and predicted abundances for the OOB and test samples. Top: The upper panel shows the [Fe/H] predictions for the OOB sample, color-coded by $g_0$. Each symbol corresponds to a galaxy: circles for Fnx, triangles for Scl, squares for UMi, and stars for Dra. Test sample results are shown in black. The number of samples and RMSE values from 5-fold CV are given at the top. The black dashed diagonal line represents the one-to-one relation. The lower panel displays residuals, with the dashed horizontal line at $\Delta$[Fe/H] = 0. Bottom: Same as the top panels, but for [Mg/H].}    
    \label{Fig:training_test_pred}
\end{figure}

\subsection{Methods}\label{sec:methods}

Using the training spectroscopic DEIMOS data, specifically, the $M_g$, $M_i$, $M_{NB515}$ magnitudes, and the color indices $(g-i)_0$, $(g-NB515)_0$, $(NB515-i)_0$ as predictor features, we trained RF regressors \citep{Breiman2001} to predict the stellar abundances of [Fe/H] and [Mg/H] separately. This trained RF was later applied to predict the [Fe/H] and [Mg/H] abundances for stars in the HSC data. Furthermore, estimates of [Mg/Fe] abundances were obtained by subtracting the [Fe/H] predictions from the [Mg/H] predictions. 

Unlike traditional statistical models which often impose restrictive assumptions (e.g., linearity), RFs are a flexible machine learning technique, capable of capturing highly complex, nonlinear relationships in the data \citep{Breiman2001, Hastie2009, Biau2015}. 
Briefly, an RF is an ensemble, or collection, of decision trees \citep{Breiman1984}. Each decision tree is constructed by recursively making binary splits (or yes-no decisions) based on some feature and threshold (e.g., $(g-i)_0 <$ 1 or $\geq1$) that are selected to maximize the reduction in impurity (or variance) and results in the most homogeneous partitions of the data at each split.
Furthermore, in the context of RFs, each decision tree is independently trained using a bootstrapped sample (i.e., sampled with replacement), and at each split, only a random subset of features are considered when determining the optimal split. This inherent randomness in RFs helps to reduce overfitting and improves its generalization performance. The final abundance estimate for each star is obtained by first computing the prediction from each decision tree --- that is, the average target value (e.g., [Fe/H] or [Mg/H]) across all training samples that fell in the same leaf (or terminal) node as the given star under study --- and then averaging these predictions across all trees in the forest.

Given that there are various hyperparameters, governing the structure of the forest, we used 5-fold cross-validation (CV) to tune the number of trees in the forest (\texttt{n\_estimators}) and the minimum number of samples required in each leaf node (\texttt{min\_samples\_leaf}). We selected the hyperparameters that minimized the cross-validated root mean squared error (RMSE), resulting in (\texttt{n\_estimators, min\_samples\_leaf}) = (500, 1) for the [Fe/H] and [Mg/H] prediction models.

To evaluate the overall RF prediction accuracy and how well the predicted abundances align with the observed abundances, we measured three different prediction metrics: 
\begin{enumerate}
    \item Root Mean Squared Error (RMSE), which measures the root of the average squared difference between the predicted and observed values and is defined as:
    \begin{align}\label{eq:rmse}
        \text{RMSE} = \sqrt{\frac{1}{N} \sum_{i=1}^{N} (y_{{\rm phot},i} - y_i)^2},
    \end{align}
    \item $R^2$, which normalizes the RMSE by the variance of the measured abundances to provide a more interpretable measure of model performance and is defined as:
    \begin{align}\label{eq:r2}
        R^2 = 1 - \frac{\sum_{i=1}^{N} (y_{{\rm phot},i} - y_i)^2}{\sum_{i=1}^{N} (y_i - \bar{y})^2},
    \end{align}
    \item Reduced $\chi^2$, which compares the model's predictions to the expected measurement error and is defined as:
    \begin{align}\label{eq:chi2}
        \text{Reduced }\chi^2 = \frac{1}{N} \sum_{i=1}^{N} \frac{(y_{{\rm phot},i}- y_i)^2}{\delta_{y_i}^2 + \sigma_{y_i}^2},
    \end{align}
\end{enumerate}

where $N$ is the number of samples (or stars), $y_{{\rm phot},i}$ is the predicted abundance for sample $i$, $y_i$ is the observed abundance for sample $i$, $\bar{y}$ denotes the mean observed abundance, $\delta_{y_i}$ is the statistical uncertainty associated with the spectroscopic measurement, and $\sigma_{y_i}$ represents the uncertainty in the predicted abundance for sample $i$. 
Generally, lower RMSE, higher $R^2$ values, and reduced $\chi^2$ values closer to 1 indicate better model performance. 
For reliable estimation of the RF generalization performance, all test prediction errors were computed using a 5-fold CV scheme.

In addition to these global prediction performance metrics, we can obtain an estimate of the RF prediction error for each star by leveraging out-of-bag (OOB) errors, as in \citet{lu2021unified}. Recall, each tree in the RF is trained using a bootstrapped sample of the training data, leaving out some samples, referred to as the OOB sample, from the training of that tree. Because these OOB sample are not used for training that particular tree, they can serve as proxies for test sample and be leveraged to provide an unbiased estimate of the prediction error between the RF's predicted abundance and the true abundance for each star. 
More specifically, for each star in the HSC sample, we first identified the leaf node containing that star in each decision tree of the RF\@. Then, for each of these leaf nodes, we computed the RMSE between the observed abundances and the tree's predicted abundances using only the OOB sample that fell into that leaf node. The estimated prediction error for each star was finally obtained by averaging these node-wise OOB RMSEs across all trees in the RF\@. The mean of these estimated prediction errors, averaged over all stars in the HSC sample, is approximately 0.289\,dex for [Fe/H] and 0.306\,dex for [Mg/H].

Meanwhile, for each star, the ``model uncertainty'' of the prediction was estimated as the standard deviation of its predicted values across all decision trees in the forest. This model uncertainty is distinct from the prediction error discussed previously. While a large prediction error indicates a large difference between the predicted and observed abundances, a large model uncertainty indicates that the forest is having trouble reaching a consensus on a star's predicted properties. For example, stars outside of the manifold of the training data have large model uncertainty. For this reason, samples with model uncertainty greater than the maximum model uncertainty observed in the training sample were excluded. 
As a result, only 80\% of the HSC sample were included, resulting in 6713 stars (Fnx: 4,967; Scl: 975; UMi: 426; Dra: 345). The matched DEIMOS-HSC sample was likewise reduced to 561 stars (Fnx: 213; Scl: 224; UMi: 54; Dra: 70). The average model uncertainty is approximately 0.275 for [Fe/H] predictions and 0.251 for [Mg/H] predictions.

There exists a challenge in determining whether the red color of a star results from higher metallicity or advanced evolutionary stage after leaving the main sequence, as both age and metallicity influence stellar colors. Consequently, younger galaxies exhibit a different color-metallicity distribution compared to older ones. To account for this, we trained a metallicity prediction model separately for Fnx, a relatively young galaxy, and for Scl, UMi, and Dra. However, these models did not outperform the one trained with all four galaxies combined. We also explored a model incorporating the established age-metallicity relationships of galaxies \citep{Weisz2014}, but it also did not enhance performance. Therefore, we proceeded with a model that trains on all galaxies together, without explicitly considering stellar ages.

\section{Results}\label{sec:results}

In this section, we assess the predictive performance of the RF model and quantify the relative importance of the input parameters. Furthermore, we present the distributions of the predicted Fe and Mg abundances through color–color and color–magnitude diagrams, as well as within the [Fe/H]–[Mg/Fe] plane.

\subsection{Model Evaluation} \label{sec:eval}

The top panel of Figure~\ref{Fig:training_test_pred} shows the OOB [Fe/H] predictions, color-coded by $g_0$, compared with the observed [Fe/H] measurements. Here, the RMSE for [Fe/H] on the OOB and test samples are both 0.265\,dex. The fact that the RMSE for the OOB sample, which are not directly used during training, is the same as that of the test sample suggests that the model is not overfitting to the training data and is capable of making valid predictions on new data. The lower panel shows the residuals between the predictions and measurements, with the dashed line indicating $\Delta \mathrm{[Fe/H]} =$ [Fe/H]$_{\rm phot}$ - [Fe/H] $= 0$. The mean $\Delta \mathrm{[Fe/H]}$ across all stars is 0.003\,dex. Among 15 OOB stars with [Fe/H] $\leq -2.5$ that are predicted to be more metal-rich than observed (7 from Scl, 4 from UMi, and 4 from Dra), the average residual is $\Delta \mathrm{[Fe/H]} = 0.740$\,dex. For the remaining stars with higher [Fe/H], the mean residual is $\Delta \mathrm{[Fe/H]} = -0.021$\,dex.
The bottom panel shows the comparison for [Mg/H], with RMSE values of 0.262\,dex and 0.263\,dex for the OOB and test samples, respectively. The overall mean of $\Delta \mathrm{[Mg/H]}$ is 0.005\,dex. Among 40 stars with [Mg/H] $\leq -1.8$ (10 from Scl, 16 from UMi, and 14 from Dra), the average is $\Delta \mathrm{[Mg/H]} = 0.347$\,dex. For 46 stars with [Mg/H] $> -0.8$, most of which are faint in $g_0$ magnitude and primarily from Fnx (43 from Fnx and 3 from Scl), the average is $\Delta \mathrm{[Mg/H]} = –0.315$\,dex. The remaining stars have an average $\Delta \mathrm{[Mg/H]}$ of 0.008\,dex. For reference, the mean of $\Delta \mathrm{[Mg/Fe]}$ is approximately 0.002\,dex.

\begin{table}[!t]
    \centering
    \caption{Random forest prediction performance}
    \hspace{-1.2cm}
    \resizebox{1.1\linewidth}{!}{
    \begin{tabular}{ccc cc}
        \toprule\toprule
        & \multicolumn{2}{c}{[Fe/H]} & \multicolumn{2}{c}{[Mg/H]}\\
        & Training & Test & Training & Test \\
        \midrule
        RMSE (dex) & 0.100 & 0.265 & 0.115 & 0.263 \\
        $R^2$ & 0.964 & 0.746 & 0.903 & 0.497 \\
        Reduced $\chi^2$ & 0.096 & 0.803 & 0.168 & 0.888 \\
        \bottomrule
    \end{tabular}
    }
    \label{tab:modelperformance}
\end{table}

\begin{figure}[!b]
    \centering
    \includegraphics[width=0.43\textwidth]{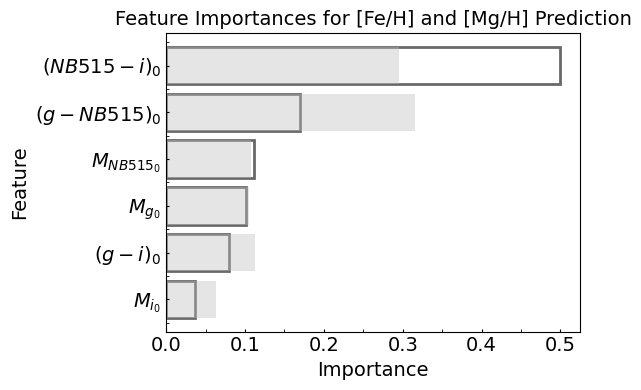}
    \caption{The feature importance used in the abundance prediction for HSC sample. The unfilled bars represent [Fe/H] predictions, while the filled bars represent [Mg/H].}
    \label{Fig:feature_importances}
\end{figure}

\begin{figure*}
    \centering
    \includegraphics[width=0.76\textwidth]{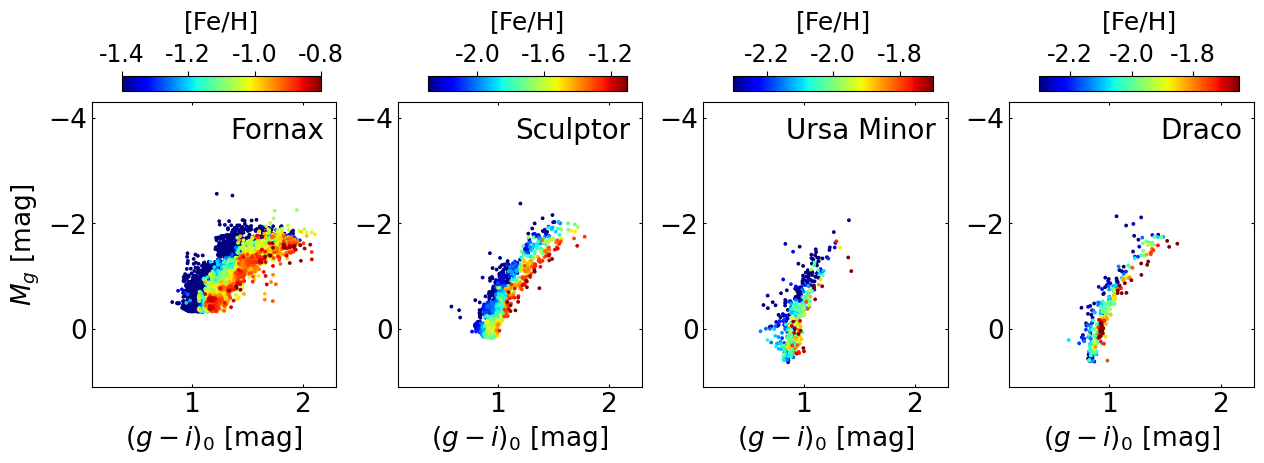}
    \caption{Color-magnitude diagrams of the HSC data with Fnx, Scl, UMi, and Dra shown from left to right, color-coded by [Fe/H] predictions.}
    \label{Fig:cmd_feh}
\end{figure*}

\begin{figure*}
    \centering
    \includegraphics[width=0.9\textwidth]{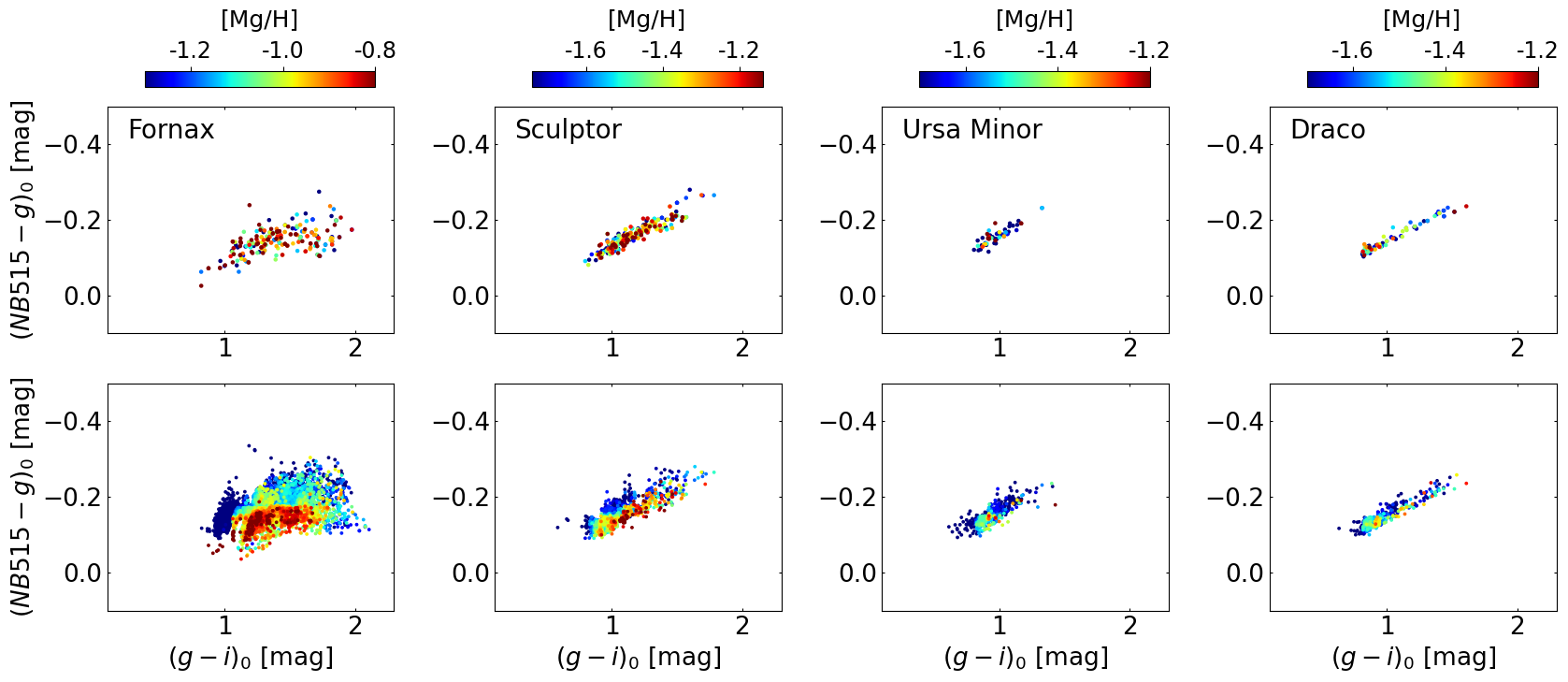}
    \caption{Top: Color-color diagrams of the DEIMOS data with Fnx, Scl, UMi, and Dra shown from left to right, color-coded by [Mg/H]. Bottom: Same as the top panel, but for [Mg/H] predictions of HSC data.}
    \label{Fig:ccd_mgh}
\end{figure*}

Overestimation of metallicity for metal-poor stars has been reported in various photometric studies. For example, \citet{Sun2025} derived photometric metallicities using RF model with broadband photometry from the Kilo-Degree Survey \citep[KiDS;][]{Kuijken2019} and the VISTA Kilo-degree Infrared Galaxy \citep[VIKING;][]{Edge2013}. Their Figure~2 shows a similar trend, which they attribute to the limited sensitivity of the broad $u$-band to metallicity. As another example, \citet{Fallows2022} applied a neural network to stars from the Two Micron All Sky Survey \citep[2MASS;][]{Skrutskie2006} and the Wide-field Infrared Survey Explorer \citep[WISE;][]{Wright2010, Schlafly2019}, and observed the similar pattern of overestimation (Figures 8 and B1). To mitigate this, they suggested increasing the number of stars at metallicity extremes as well as incorporating bluer narrow-band filters. They also proposed that the underestimation on the metal-rich side may reflect the presence of intermediate-age stars, as proposed by \citet{Lianou2011} based on photometric data from \citet{Coleman2005} and \citet{Walker2007, Walker2009}.
While [Fe/H] can be reasonably predicted from observations in the $g$-band and redder wavelengths when combined with isochrone-based approaches \citep[e.g.,][]{Ogami2025}, enhancing metallicity sensitivity in the $u$-band can improve purely data-driven predictions.

Recent large-scale photometric surveys such as the Javalambre-Physics of the Accelerated Universe Astrophysical Survey \citep[J-PAS;][]{Benitez2014}, the Javalambre Photometric Local Universe Survey \citep[J-PLUS;][]{Cenarro2019}, and the Southern Photometric Local Universe Survey \citep[S-PLUS;][]{MendesdeOliveira2019} address this limitation by incorporating a set of seven narrow-band filters, among which one targets the metallicity-sensitive \ion{Ca}{2} H\& K absorption features at 3933.7 and 3968.5~\AA, and another targets the Mg b triplet. These efforts have enabled accurate estimation of both [Fe/H] and [Mg/Fe], as demonstrated in studies such as \citet{Yang2022}, \citet{Huang2024}, and \citet{FerreiraLopes2025}. For instance, \citet{Huang2024} employed kernel principal component analysis \citep[KPCA;][]{scholkopf1998} to estimate these abundances from J-PLUS DR3 data, reporting a median offset and standard deviation of approximately –0.02\,dex and 0.11\,dex for [Fe/H] (in the range –3.5 $<$ [Fe/H]), and +0.01\,dex and 0.08\,dex for [Mg/Fe] (–0.2 $<$ [Mg/Fe] $\leq$ 0.6), respectively (see their Figure~9). Within the same abundance ranges, our residuals yield median and standard deviation values of –0.023\,dex and 0.304\,dex for [Fe/H], and +0.030 and 0.190\,dex for [Mg/Fe].
These findings suggest that incorporating the existing $NB395$ narrow-band filter, along with medium-band filters currently under development for improved metallicity calibration in HSC\footnote{\url{https://sites.google.com/view/hsc-mb-workshop/home?authuser=0}}, could lead to more precise abundance estimates with reduced systematic offsets in future studies.

In addition, we found that the predicted abundance range is somewhat narrower than the spectroscopic range for both [Fe/H] and [Mg/H], due to over- and under-estimations near the ends of the distribution, with this trend being slightly more pronounced for [Mg/H]\@. We suspect that this is partly attributable to the intrinsically narrower distribution of [Mg/H] in the training sample, which makes it inherently more challenging to model compared to [Fe/H]\@. In addition, regression methods such as RF are trained to learn conditional mean (i.e., average), and thus tend to avoid extreme predictions near the boundaries of the training distribution.
In conclusion, future work should consider incorporating isochrones that account for stellar ages, employing a multiband photometric system including metallicity-sensitive filters, exploring alternative machine learning models, and most importantly, securing additional spectroscopic abundance measurements. These improvements will enhance model accuracy and allow the predicted abundance distributions to better match those of spectroscopic data.

Furthermore, the training and test RMSE (Eq.~\ref{eq:rmse}), $R^2$ (Eq.~\ref{eq:r2}), and reduced $\chi^2$ (Eq.~\ref{eq:chi2}) values, obtained from 5-fold CV, are shown in Table~\ref{tab:modelperformance}. The 0.100 and 0.115 RMSE values for the training sample for [Fe/H] and [Mg/H] are lower than those for the test sample 0.265 and 0.263, which aligns with the expectation that the model tends to perform better on data it has been trained on.
The $R^2$ values for the [Fe/H] and [Mg/H] training sample are 0.964 and 0.903, respectively, indicating that the model achieves high accuracy in prediction. For the test sample, the corresponding values are 0.746 and 0.497.
Similarly, the $\chi^2$ values for the [Fe/H] and [Mg/H] training sample are 0.096 and 0.168, respectively, further confirming that the model predictions closely follow the true values when applied to the training data. For the test sample, the $\chi^2$ values are 0.803 and 0.888, which are relatively close to 1, suggesting that despite minor discrepancies, the model predictions generally fall within the expected error range.

\subsection{Feature Importances} \label{sec:importance}

Beyond prediction, we quantified the importance of each predictor in the RF model using the mean decrease in impurity (MDI) feature importance score \citep{Breiman2001}. A higher MDI indicates that the feature is more important in making the model's predictions.
We summarize the feature importance results in Figure~\ref{Fig:feature_importances}, where unfilled and filled bars represent the importance in the [Fe/H] and [Mg/H] prediction models, respectively. 
For [Fe/H] predictions, we found that the color index $(NB515-i)_0$ was the most important feature, contributing 50.0\% of the overall importance. This is followed by $(g-NB515)_0$, $M_{NB515_{0}}$, $M_{g_{0}}$, $(g-i)_0$, and $M_{i_{0}}$, which contributed 17.0\%, 11.2\%, 10.2\%, 7.9\%, and 3.7\%, respectively. This suggests that color indices involving both the narrow and broadband filters, especially those spanning a wide wavelength range, such as $i$ and $NB515$, are important for predicting [Fe/H]\@.
For [Mg/H], the most important feature was $(g-NB515)_0$, followed closely by $(NB515-i)_0$, contributing 31.6\% and 29.6\%, respectively. The remaining contributions came from $(g-i)_0$, $M_{NB515_{0}}$, $M_{g_{0}}$, and $M_{i_{0}}$, at 11.2\%, 10.8\%, 10.5\%, and 6.3\%, respectively. This indicates that, similar to [Fe/H], both color indices involving the $NB515$ filter and the absolute magnitudes play substantial roles in prediction [Mg/H]\@. However, the more distributed importance values for [Mg/H] suggest that a balanced use of multiple photometric features may enhance prediction stability.

\subsection{Color-Magnitude and Color-Color Diagram}

Figure~\ref{Fig:cmd_feh} shows a color–magnitude diagram, color-coded with the [Fe/H] predictions for the HSC data. Similar to Figure~\ref{Fig:tailoredmag}, the trend from metal-rich to metal-poor stars is clearly observed from the right to the left. Assuming that the members of each dwarf galaxy have the same age, it is expected that metal-poor stars would appear in the bluer region. However, since younger stars that formed more recently may still appear hot and blue, this necessitates future comparisons with isochrone models. Additionally, small segmented patterns may appear in certain areas, which can be attributed to the binary choices made by the RF model at every branch. While smoothing could be achieved by further tuning hyperparameters, we opted to retain the hyperparameter values that minimize the RMSE\@.

Figure~\ref{Fig:ccd_mgh} shows the [Mg/H] predictions on the color-color diagram. As expected, we see a clear trend of decreasing [Mg/H] values from the lower right to the upper left. If we assume the stars have the same age, the Mg-poor stars, marked by the blue dots, are likely to have been formed during the early stages of the galaxy's chemical evolution. However, contamination from foreground MSTO stars with higher effective temperatures and lower color indices may still be present even after the $NB515$-based membership cut. Furthermore, even at fixed metallicity, variations in age can lead to differences in temperature and luminosity, potentially shifting stellar positions on the color–color diagram. Future work can improve on our analysis by accounting for stellar temperature, age, and evolutionary stage.

\subsection{[Fe/H], [Mg/H], and [Mg/Fe] Predictions}

\begin{figure*}
    \centering
    \includegraphics[width=.9\textwidth]{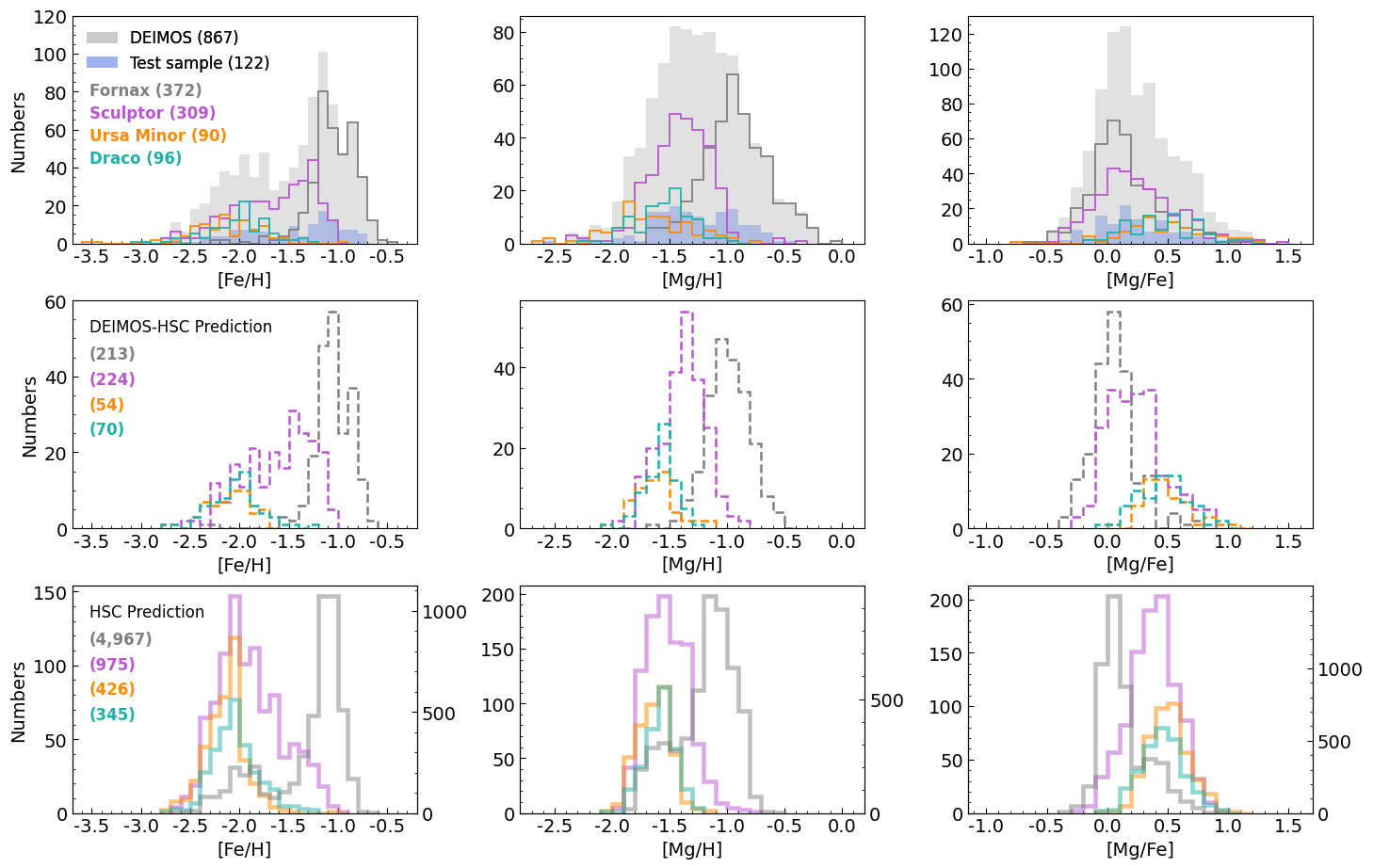}
    \caption{Top: The distribution of [Fe/H], [Mg/H], and [Mg/Fe] for the DEIMOS sample, shown from left to right. The combined distribution of all DEIMOS samples is shown as a filled gray histogram. The filled blue histogram represents the abundance prediction results of the test sample. Colors gray, purple, orange, and green correspond to the spectral measurements for Fnx, Scl, UMi, and Dra, respectively. The number of samples for each is listed on the left. Middle: Similar to the top, but showing prediction results for the DEIMOS-HSC sample as dashed lines. Bottom: Similar to the middle, but for the HSC sample.}
    \label{Fig:compare_feh_mgh_mgfe}
\end{figure*}

The top panels of Figure~\ref{Fig:compare_feh_mgh_mgfe} show the distributions of [Fe/H], [Mg/H], and [Mg/Fe] for the DEIMOS sample. The prediction results of the test sample closely follow the abundance distributions of the DEIMOS sample. The middle panels show the predicted abundance distributions for the DEIMOS-HSC sample. While the RF model’s tendency to regress toward the mean can affect the recovery of the full range, the overall shape of the distributions are generally consistent with those of the DEIMOS sample shown above. The bottom panels present the predicted abundance distributions from HSC data for the four galaxies. The mean and standard deviation of the predicted abundance distributions are summarized in Table~\ref{tab:abundance_mean_std}. 

\begin{table*}
    \centering
    \caption{Mean and standard deviations of [Fe/H], [Mg/H], and [Mg/Fe]}
    \hspace{-1.6cm}
    \resizebox{1.05\textwidth}{!}{%
    \begin{tabular}{l ccc ccc ccc}
        \toprule\toprule
        & \multicolumn{3}{c}{[Fe/H]} 
        & \multicolumn{3}{c}{[Mg/H]} 
        & \multicolumn{3}{c}{[Mg/Fe]} \\
        \cmidrule(lr){2-4} \cmidrule(lr){5-7} \cmidrule(lr){8-10}
        & DEIMOS & DEIMOS-HSC & HSC 
        & DEIMOS & DEIMOS-HSC & HSC 
        & DEIMOS & DEIMOS-HSC & HSC \\
        \midrule
        Fnx & -1.046 $\pm$ 0.248 & -1.045 $\pm$ 0.183 & -1.323 $\pm$ 0.396
            & -0.920 $\pm$ 0.295 & -0.991 $\pm$ 0.180 & -1.193 $\pm$ 0.253
            &  0.125 $\pm$ 0.280 &  0.054 $\pm$ 0.181 &  0.130 $\pm$ 0.195 \\
        Scl & -1.647 $\pm$ 0.414 & -1.616 $\pm$ 0.349 & -1.895 $\pm$ 0.330
            & -1.409 $\pm$ 0.275 & -1.384 $\pm$ 0.194 & -1.521 $\pm$ 0.187
            &  0.238 $\pm$ 0.325 &  0.232 $\pm$ 0.235 &  0.374 $\pm$ 0.201 \\
        UMi & -2.140 $\pm$ 0.382 & -2.103 $\pm$ 0.227 & -2.141 $\pm$ 0.203
            & -1.665 $\pm$ 0.358 & -1.615 $\pm$ 0.180 & -1.636 $\pm$ 0.143
            &  0.475 $\pm$ 0.343 &  0.488 $\pm$ 0.189 &  0.504 $\pm$ 0.164 \\
        Dra & -1.996 $\pm$ 0.324 & -2.041 $\pm$ 0.263 & -2.043 $\pm$ 0.239
            & -1.542 $\pm$ 0.241 & -1.579 $\pm$ 0.138 & -1.586 $\pm$ 0.132
            &  0.454 $\pm$ 0.296 &  0.462 $\pm$ 0.205 &  0.456 $\pm$ 0.179 \\
        \bottomrule
    \end{tabular}%
    }
    \label{tab:abundance_mean_std}
\end{table*}

\begin{figure*}
    \centering
    \includegraphics[width=0.85\textwidth]{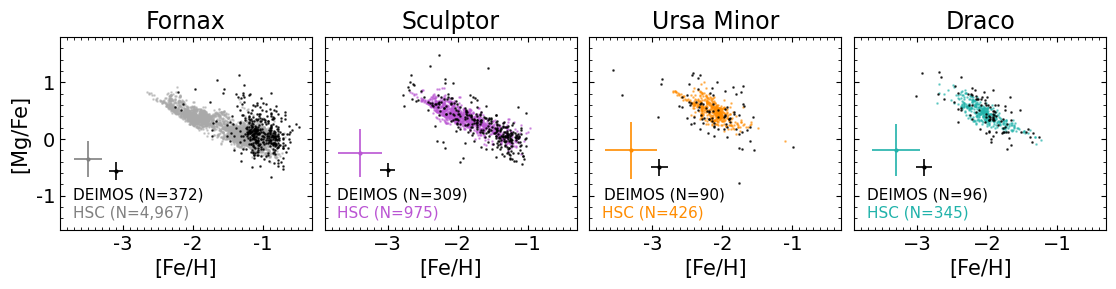}
    \caption{The [Fe/H] vs.\ [Mg/Fe] distributions for the DEIMOS and HSC samples. From left to right, the panels correspond to Fnx, Scl, UMi, and Dra. Black points represent the DEIMOS measurements, and black error bars indicate their representative mean errors derived from statistical uncertainties. Gray, purple, orange, and green points in each panel indicate the photometric predictions from the HSC sample, with error bars showing their corresponding mean error (from the OOB method discussion in Sec.~\ref{sec:methods}). The number of stars in each dataset is shown in the bottom-left corner of each panel.}
    \label{Fig:spec_phot_mgfe_feh}
\end{figure*}

Similar to the previous spectroscopic results \citep{Battaglia2006, Starkenburg2010, Kirby2011b, Hendricks2014, Hasselquist2021}, the photometric metallicity distribution function (MDF) of Fnx, shown in gray, gradually rises and exhibits a steep decline after peaking at [Fe/H] = -1.1. The distribution of [Mg/H] follows a comparable trend, but with a peak at [Mg/H] = -1.2. The distributions of [Mg/Fe] for all galaxies, including Fnx, are derived from the corresponding [Mg/H] and [Fe/H] prediction values. Galaxy properties inferred from these distributions, such as SFHs, are discussed in Section~\ref{sec:discussion}. 
The MDF of Scl, shown in purple for both the DEIMOS and DEIMOS-HSC samples, exhibits a bimodal structure consistent with the results of \citet{Kirby2011b} and \citet{delosReyes2022}. Our median value of ${\rm [Fe/H]} = –1.9 \pm 0.3$ also is in good agreement with \citet{Barbosa2025}. However, the MDF prediction for the HSC sample shows a more prominent peak at ${\rm [Fe/H]} = -2.1$, followed by a smaller secondary peak at ${\rm [Fe/H]} = -1.5$. This is likely attributable to the inclusion of a larger number of stars located in the outer regions of the galaxy, which harbors metal-poor stars, in our photometric sample. The [Mg/H] distribution shows a distinct Mg-enhanced tail following the peak at ${\rm [Mg/H]} = –1.6$.
The [Fe/H] distribution of UMi, shown in orange, exhibits the most prominent peak among the four galaxies at ${\rm [Fe/H]} = –2.1$, and a slight metal-rich tail, which is also seen in spectroscopy. 
Dra contains slightly more stars in its metal-rich tail compared to UMi, resulting in a mean that is about 0.1\,dex higher than those of UMi. The [Mg/H] distributions of the UMi and Dra are quite similar, both showing a peak at around $-1.6$ and comparable standard deviations.

Figure~\ref{Fig:spec_phot_mgfe_feh} shows [Mg/Fe] as a function of [Fe/H] for DEIMOS and HSC\@. From left to right, the panels represent Fnx, Scl, UMi, and Dra. The black points represent the DEIMOS spectroscopic measurements, with their mean error values indicated by the black error bars, and the total number of stars shown at the bottom of each panel. The colored points in each panel correspond to the HSC photometric predictions for each galaxy, with their respective mean errors and sample sizes also indicated. 
In general, the predicted abundance patterns of increasing [Mg/Fe] with decreasing [Fe/H] broadly agrees with the DEIMOS sample and previous spectroscopic studies: for example, \citet{Letarte2010}, \citet{Lemasle2014}, \citet{Reichert2020} for Fnx, \citet{Letarte2010}, \citet{Hill2019}, \citet{Reichert2020} for Scl, \citet{Cohen2010}, \citet{Reichert2020} for UMi, and \citet{Shetrone2001}, \citet{Cohen2009} for Dra. While the predicted range of [Mg/Fe] at fixed [Fe/H] is somewhat narrower than that of the spectroscopic training samples across all galaxies, and the photometric predictions for UMi and Dra do not fully extend to the most metal-poor stars, these truncated distributions likely reflect both the limitations of the model and the slight overestimation of metal-poor stars discussed in Section~\ref{sec:eval}. The median [Fe/H] vs. [Mg/Fe] patterns and their spatial variations are discussed in more detail in the following section.

\begin{figure*}
    \centering
    \includegraphics[width=0.9\textwidth]{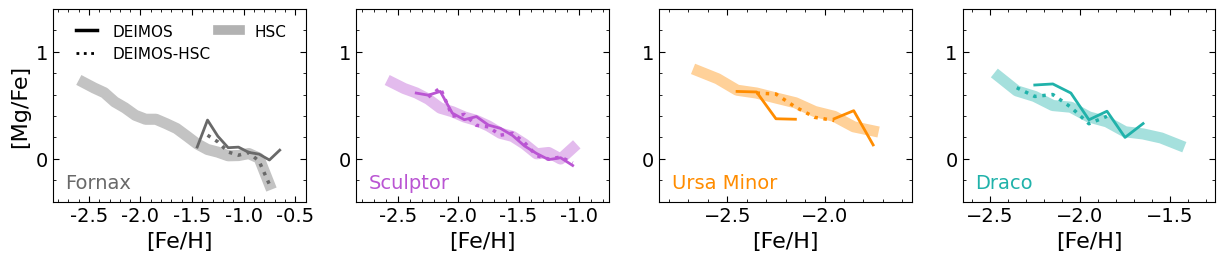}
    \caption{Observed trends of [Mg/Fe] vs.\ [Fe/H] for spectroscopy (DEIMOS, thin solid line) and photometry (HSC, thick bars). The trend of the predictions for DEIMOS-HSC stars in the DEIMOS spatial footprint is indicated by the dashed line. The samples are divided into 0.1\,dex bins of [Fe/H] and [Mg/Fe], and the lines connecting the bins represent the median values for bins containing at least five stars. Gray, purple, orange, and green indicate Fnx, Scl, UMi, and Dra from left to right, respectively.}
    \label{Fig:spec_phot_mgfe_feh_median}
\end{figure*}

\begin{figure*}
    \centering
    \includegraphics[width=0.95\textwidth]{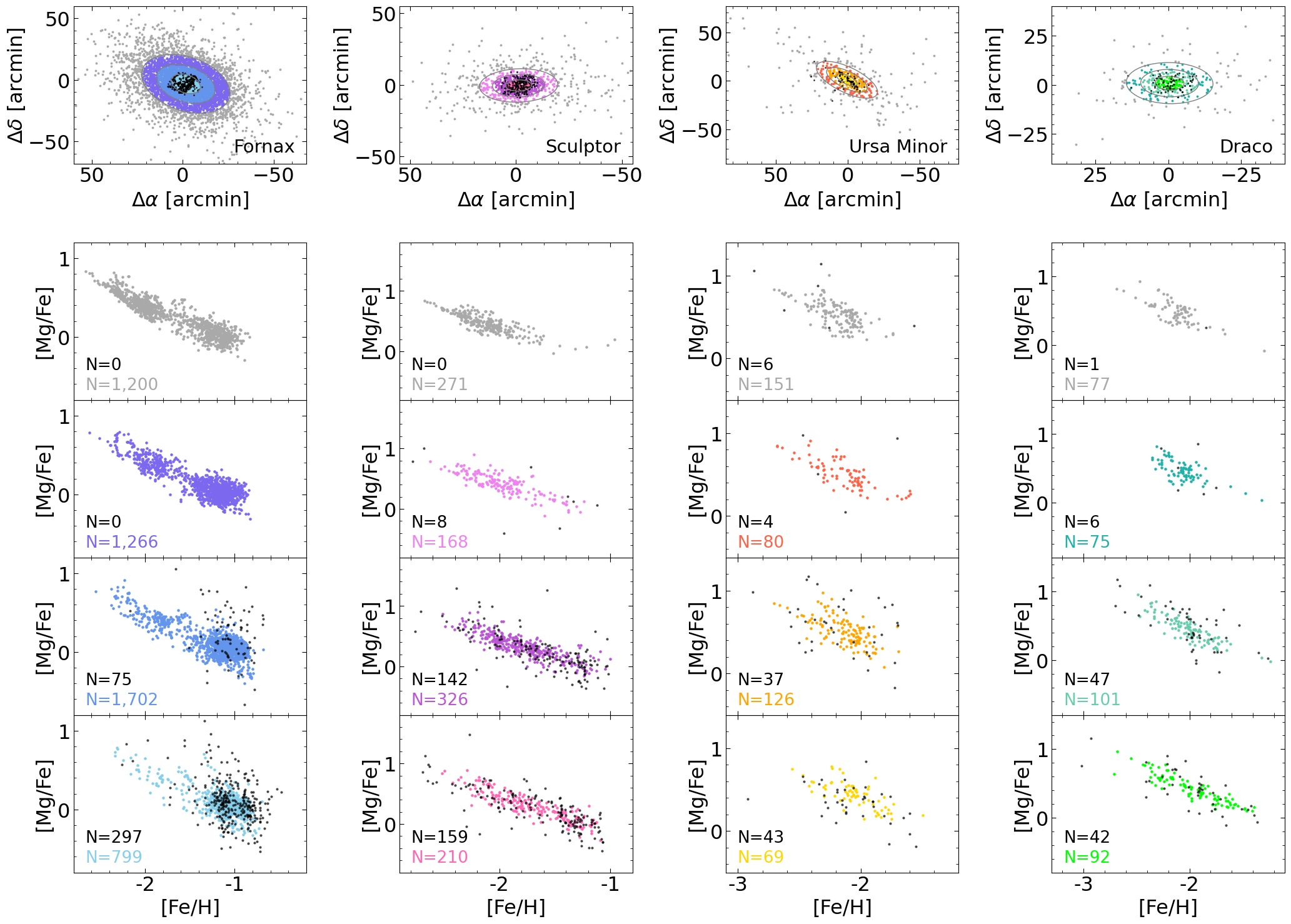}
    \caption{Top row: The spatial distributions of stars in Fnx, Scl, UMi, and Dra are shown from left to right. In each panel, four color-coded radial regions extend from the galaxy center to the outskirts. Each region corresponds to 0.5, 1.0, and 1.5 times the half-light radius $r_h$ of each galaxy, adopted from \citet{McConnachie2020}. The innermost black dots indicate the DEIMOS spectroscopic sample. Second to fifth rows: The corresponding spatial distributions of [Fe/H] and [Mg/Fe] for each radial region, arranged from the outermost (top) to the innermost regions (bottom). The number of stars in each DEIMOS and HSC sample is shown in the bottom-left corner of each panel, with DEIMOS on the top line. The black dots denote the DEIMOS sample.}
    \label{Fig:spec_phot_mgfe_feh_radec}
\end{figure*}

\begin{figure*}
    \centering
    \includegraphics[width=0.45\textwidth]{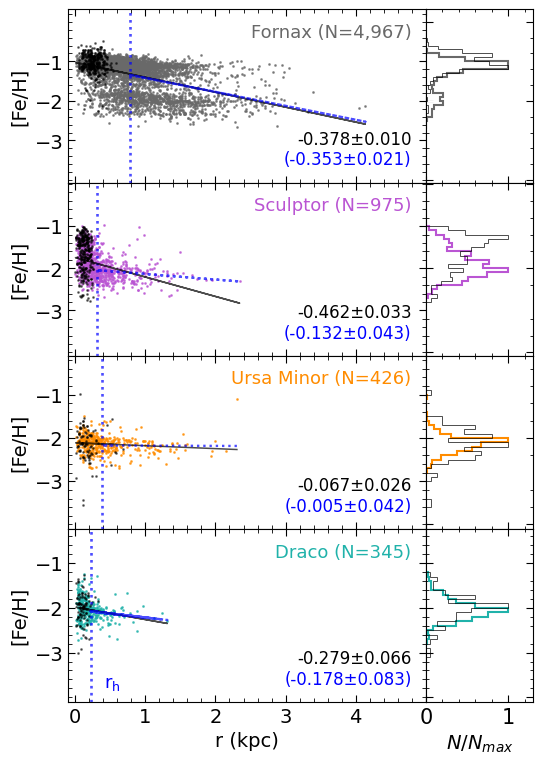}
    \includegraphics[width=0.45\textwidth]{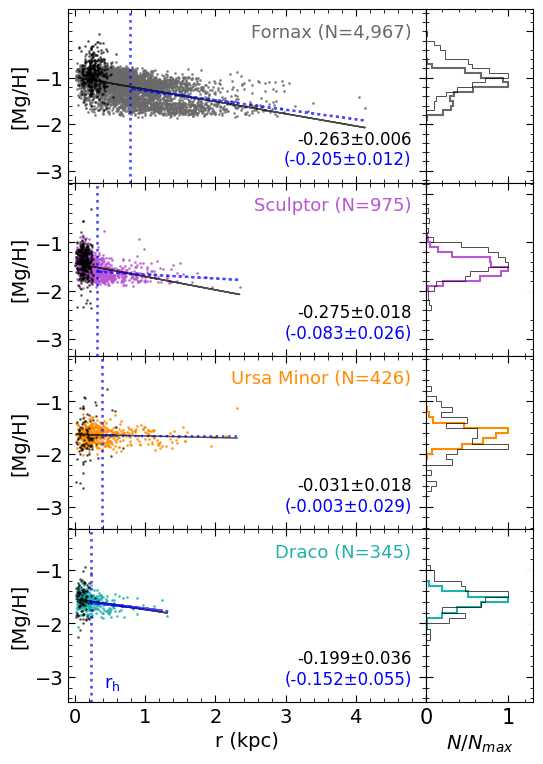}
    \caption{Left: Radial distributions of the [Fe/H] predictions for the HSC sample. The black dots represent the DEIMOS sample. From top to bottom, the panels correspond to Fnx, Scl, UMi, and Dra, with the number of stars indicated in the top-right corner of each panel. The vertical dashed blue line marks the $r_h$, adopted from \citet{McConnachie2020}. The least-squares linear fits are shown as the solid black and dashed blue lines, measured from the galaxy center and from the $r_h$ outward, respectively. The corresponding slopes are displayed in the bottom-right corner of each panel, with those derived beyond $r_h$ (dotted blue lines) shown in the line below. The histograms on the right are normalized to their peak values, and the black histograms represent the DEIMOS sample. Right: Same as left panels, but for [Mg/H].}
    \label{Fig:spatialdistribution}
\end{figure*}

\begin{figure*}
    \centering
    \includegraphics[width=0.45\textwidth]{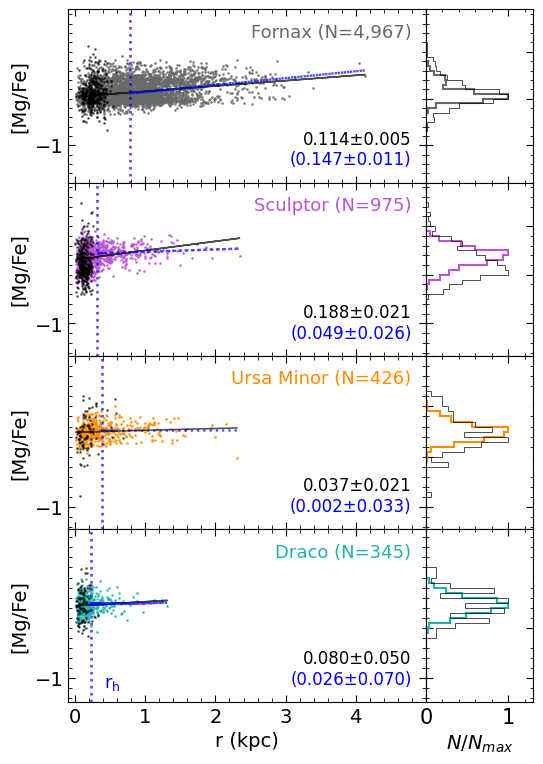}
    \caption{Same as Figure~\ref{Fig:spatialdistribution}, but for [Mg/Fe].}
    \label{Fig:spatialdistribution_mgfe}
\end{figure*}

\section{Discussion}\label{sec:discussion}

Building on the work of \citet{Lianou2011}, which estimated [Fe/H] from photometric data in dwarf galaxies, the next step is to leverage the predicted elemental abundances from larger photometric samples to explore abundance patterns in regions beyond the reach of spectroscopy and to infer the SFHs of dwarf galaxies.
For example, Figure~\ref{Fig:spec_phot_mgfe_feh_median} illustrates that the photometric predictions extend beyond the spectroscopically covered regions. Considering the uncertainties in both spectroscopic and photometric data, we found that the photometric predictions beyond the dashed lines---and therefore beyond the areal coverage of DEIMOS---continue to follow similar trends. 

In more detail, for Fnx, the photometric [Mg/Fe] predictions gradually decline from [Fe/H] = –2.5 and show a slight $\alpha$-knee around [Fe/H] = –1.9 with [Mg/Fe] = +0.4, consistent with \citet{Hendricks2014}. Then a mild bump appears at [Fe/H] = –0.9, followed by a decline toward –0.7, in agreement with \citet{Letarte2010}. The DEIMOS–HSC estimates follow the spectroscopic trend up to [Fe/H] = –1.2, beyond which they align more closely with the photometric predictions. The DEIMOS spectroscopic [Mg/Fe] also shows a declining trend with increasing [Fe/H], but exhibits a peak near [Fe/H] = –1.4, likely due to a few outliers. These features suggest a gradual increase in Type Ia SNe contributions within Fnx.

For Scl, the [Mg/Fe] predictions gradually decline from about +0.7 at [Fe/H] = –2.5, consistent with \citet{Kirby2011a}. This smooth decrease without a clear $\alpha$-knee may indicate that Type Ia SNe started enriching Scl at an early stage \citep{Kirby2011a}, or it may result from our model’s tendency to predict values near the mean, influenced by the scatter in samples at [Fe/H] $\leq$ –1.5. Around [Fe/H] = –1.25, a slight [Mg/Fe] peak is observed, followed by a dip and a rise to about +0.2\,dex near [Fe/H] = –1.15. As shown in the high-resolution EAGLE simulations by \citet{Mason2024}, the absence of a distinct $\alpha$-knee may reflect a sustained balance between CCSNe and Type Ia SNe contributions, consistent with prolonged steady SF, while the upturn at the metal-rich end may indicate recent SF activity. Meanwhile, the DEIMOS sample exhibits peaks near [Fe/H] = –2.1 and –1.8 \citep{Hill2019}, as well as a smaller peak near –1.15, with DEIMOS-HSC estimates generally following these trends. Considering Figure~\ref{Fig:spec_phot_mgfe_feh}, the scattered [Mg/Fe] values among metal-poor stars cast doubt on the photometric peak at –2.1 and on the [Mg/Fe] $\approx$ +0.8 values at [Fe/H] = –2.8, which are about 0.3\,dex higher than reported by \citet{Skuladottir2021, Skuladottir2024}. Therefore, a larger sample of metal-poor stars is necessary to clarify these features \citep{deBoer2012a}.

For both UMi and Dra, the photometric [Mg/Fe] trends show a smooth decline with increasing [Fe/H], without prominent peaks, and the DEIMOS-HSC estimates generally follow similar behavior. In UMi, the $\alpha$-knee has been identified near [Fe/H] = –2.1 to –2.2, and stars with [Fe/H] $\leq$ –2.5 have been observed to reach [Mg/Fe] = +0.5 \citep{Reichert2020, Sestito2023}. However, our spectroscopic trend appears truncated in this regime due to fewer than five stars per bin. Additional medium- and high-resolution spectra, such as from the ongoing Subaru Prime Focus Spectrograph (PFS) survey \citep{Takada2014}, will be essential to reduce scatter and better constrain the trend at [Fe/H] $\leq$ –2.
In Dra, the spectroscopic trend shows a broad peak near [Fe/H] = –2.1, followed by a decline and a minor rise of about +0.1\,dex near –1.95 \citep{Venn2014} and –1.75, broadly reproduced by the DEIMOS-HSC predictions. In contrast, its photometric trend declines monotonically, as in UMi, possibly reflecting a more smoothly decreasing SF rate. Dra's $\alpha$-knee is expected to occur at the lowest metallicity among the four galaxies, around [Fe/H] = –2.7 \citep{Hendricks2014}, but due to a lack of samples below [Fe/H] = –2.5, neither spectroscopic nor photometric trends can be evaluated in this regime. Further observations of metal-poor stars are needed to better constrain the trend at low metallicities.

Figure~\ref{Fig:spec_phot_mgfe_feh_radec} shows a more detailed distribution of [Fe/H] vs.\ [Mg/Fe] for different radial regions of the galaxies. From these distributions, we observe an increasing fraction of metal-poor \citep{Battaglia2006, deBoer2012a, deBoer2012b, Tolstoy2023} and Mg-rich stars toward the outskirts. This highlights the key advantage of the enlarged photometric samples, which enables the study of abundance patterns in the galaxy outskirts—regions that are underrepresented or inaccessible in the spectroscopic samples, which is predominantly concentrated in the central regions. Another significant feature is concentration of [Mg/Fe] predictions around [Fe/H] = –1 in Fnx, which deviates from the monotonically decreasing trends observed in the other three galaxies. This supports previous studies reporting that Fnx experienced an extended SFH, continuing into more recent times, while the other galaxies ceased SF after an early phase.

Figures~\ref{Fig:spatialdistribution} and \ref{Fig:spatialdistribution_mgfe} present the radial abundance gradients of [Fe/H], [Mg/H], and [Mg/Fe] predictions as a function of galactocentric radius. The photometric predictions extend to radii approximately 7, 9, 3, and 4 times the spectroscopically measured half-light radius $r_h$ for Fnx, Scl, UMi, and Dra, respectively. These gradients provide insight into how SF has changed with distance from the galaxy center. The decreasing [Fe/H] predictions gradient of Fnx is consistent with the results of \cite{Walker2009} and \cite{Kirby2011a}, but appears significantly steeper, likely due to the abundance of [Fe/H] predictions below –1.8 in our analysis. Additionally, the slope from $r_h$ to the outer region of the galaxy does not differ significantly from the slope measured from the center outward. This suggests that SF gradually became more centrally concentrated, although it may have occurred more uniformly throughout the galaxy at earlier times. However, it should be noted that the photometric [Fe/H] values peaking around $–2$ in this region lie beyond the coverage of the DEIMOS sample and may therefore result from extrapolation. Thus, it may simply reflect the presence of stars with [Fe/H] $<$ –1.8 in the outskirts, rather than indicating a true concentration of metal-poor stars.

In contrast, Scl shows a comparable distribution range between the DEIMOS and HSC samples in terms of both magnitude and color (Figure~\ref{Fig:tailoredmag}). The negative [Fe/H] gradient from the galaxy center toward the outer regions aligns with the results of previous studies \citep{Walker2009, Kirby2011b, Tolstoy2023, Barbosa2025}. Beyond $r_h$, however, the gradient becomes noticeably shallower, a pattern also reported by \citet{Barbosa2025}, although our measured slope is shallower in comparison. This aligns with the presence of a more distinctly metal-poor center \citep{Bettinelli2019}, a higher concentration of metal-rich stars near the center, and a less dispersed [Mg/H] gradient compared to Fnx. These results may reflect the scenario proposed by \citet{Bettinelli2019}, in which a short and intense period of SF occurred in the peripheral regions, followed by a more prolonged episode concentrated in the central region. 

UMi, which exhibits a very slight decreasing [Fe/H] gradient, is consistent with recent studies \citep{Kirby2011b, Jin2016, Pace2020, Sestito2023}. Beyond $r_h$, the [Fe/H] and [Mg/H] gradient becomes even flatter. Although the limited number of predictions at the low and high ends of the Fe and Mg abundance distributions may lead to some variation in the estimated slope in future studies, the current results suggest a distinctly shallow gradient. According to the analysis by \citet{Kirby2011b}, this may be attributed to UMi’s very short SF duration. Alternatively, the gradient can also be understood in terms of its dynamical properties. As suggested by \citet{Martinez-Delgado2001} and \citet{Jin2016}, tidal forces between UMi and the Milky Way may have stirred the stellar component, resulting in a well-mixed [Fe/H] distribution and consequently a weak gradient. This is also in line with the 3D hydrodynamic simulations by \citet{Marcolini2006}, which indicate that external mechanisms, such as ram pressure stripping or tidal interactions with the host galaxy, may have significantly influenced the galaxy's evolution. UMi also exhibits the smallest gradients in both [Mg/H] and [Mg/Fe].

Dra, which has similar luminosity to UMi as well as a comparable orbit around the Milky Way \citep{Miyoshi2020, Pace2022}, exhibits a decreasing [Fe/H] gradient, in agreement with previous studies \citep[e.g.,][]{Kirby2011b, Jin2016, Han2020}. The overall slope falls between those of UMi and Scl, while the outer ($r > r_h$) gradient is slightly steeper than that observed in Scl. This suggests that Dra likely experienced a more extended and gradual decline in SF compared to UMi, but a more rapid cessation than in Scl \citep{Kirby2011a}. Consistent with our findings, high-resolution cosmological zoom-in simulations of Milky Way–like galaxies by \citet{Hirai2024} showed [Fe/H] and [Mg/Fe] gradients that lie between our predicted gradients for UMi and Dra. Furthermore, their results suggest that such systems likely formed more than 13 Gyr ago, allowing us to infer that Dra may likewise be of similarly ancient origin. With the addition of future spectroscopic samples and the application of multi-zone chemical evolution modeling, it will become possible to better constrain the duration and intensity of SF episodes, thereby shedding light on the detailed SFHs of these dwarf galaxies.

\section{Summary}\label{sec:summary}

To better understand the chemical abundance distributions of the Fnx, Scl, UMi, and Dra dwarf galaxies, we computed photometric estimates of Fe and Mg abundances for stars observed with Subaru/HSC\@. The HSC photometric survey utilized broad $g$ and $i$ filters alongside a narrow $NB515$ filter, enabling the distinction between member giant stars and foreground dwarf stars, as well as estimation of Mg abundances from the \ion{Mg}{1}~b triplet near 5150 \AA\@. After applying photometric uncertainty cuts, Gaia DR3 proper motion filtering, and membership selection, we selected 9371 stars for the analysis. For the training set, we used a spectroscopic sample of 610 stars from medium-resolution (R $\sim$ 6000) Keck/DEIMOS spectra, matched to HSC sources within 1'', with the same selection criteria applied, using [Fe/H] and [Mg/H] as labels.

We employed RF to predict [Fe/H] and [Mg/H] abundances from HSC photometry, using three absolute magnitudes and three color indices as input features. The [Mg/Fe] ratio was derived from the predicted [Mg/H] and [Fe/H] values. RF is an ensemble method well suited for capturing complex nonlinear relationships. Each decision tree is trained on a random subset of the training data, and predictions are obtained by averaging the outputs from all trees. Model accuracy and the final prediction error for each HSC sample were evaluated using the RMSE of OOB sample, which are training stars not used in the construction of a given tree. This resulted in average errors of approximately 0.289 for [Fe/H] and 0.306 for [Mg/H]. The model uncertainty was defined as the standard deviation across all trees, with mean values of about 0.275 for [Fe/H] and 0.251 for [Mg/H]. The hyperparameters were tuned to minimize the RMSE on the test sample. Feature importance analysis revealed that the $(NB515-i)_0$ and $(g-NB515)_0$ color indices contributed most significantly to the prediction of both [Fe/H] and [Mg/H]. 

The color-magnitude and color-color diagrams of the HSC data, limited to the 80\% of stars with the smallest uncertainties, led to a final sample of 6713 stars, more ten times larger than our spectroscopic dataset. When compared with accurate abundance measurements from DEIMOS, the prediction results showed a ``rainbow'' pattern in both diagrams, consistent with expectations if all stars in a dwarf galaxy shared the same age. Furthermore, by analyzing the [Fe/H] vs.\ [Mg/Fe] predictions across the central and outer regions of each galaxy, we confirmed the [Fe/H] and [Mg/Fe] distribution pattern in the central regions, as previously identified in spectroscopic studies. Additionally, we traced the negative abundance gradients ($d{\rm [Fe/H]}/dr$ and $d{\rm [Mg/H]}/dr$) to larger radii than have been probed before in these dwarf spheroidal satellite galaxies (dSphs).  

From the chemical abundance distribution, we reaffirmed and inferred the chemical enrichment and SFHs of the four dwarf galaxies, likely undergoing early SF that enriched their surroundings with magnesium and iron from CCSNe, followed by more centrally concentrated, prolonged SF that preferentially increased their iron content. However, to understand the burstiness, efficiency, environmental influence, and the unfolding of multiple SF episodes, chemical evolution modeling is essential. Future work will also include more precise stellar ages through isochrone fitting. We also anticipate that additional medium- and high-resolution spectroscopic data, especially at the metal-poor end, will enhance the precision of abundance predictions. The ongoing Subaru PFS survey \citep{Takada2014} will be an excellent source of an expanded training set. With this growing training set, future studies could incorporate additional narrow-band filters, such as $NB395$ and $NB430$, available with HSC to refine iron-abundance estimates using \ion{Ca}{2} H\& K and to evaluate the potential of constraining carbon abundances from the CH $G$-band. In addition, ongoing large-scale multiband photometric surveys, such as J-PAS \citep{Benitez2014}, J-PLUS \citep{Cenarro2019}, and S-PLUS \citep{MendesdeOliveira2019} along with other machine learning techniques, are expected to enable the prediction of additional elemental abundances.  These advances will contribute to a deeper understanding of the evolutionary histories of the Milky Way's dwarf galaxies.

\begin{acknowledgements}
The authors express thanks to Alex Dowling, Ke Wang, Christopher Sweet, Charles Vardeman, Rosemary Wyse, Carrie Fillion, and Keyi Ding for their input.
The authors also thank the anonymous referee for insightful suggestions that enhanced the quality of this paper.

This research is based in part on data collected at the Subaru Telescope, which is operated by the National Astronomical Observatory of Japan. 
Some of the data presented herein were obtained at Keck Observatory, which is a private 501(c)3 non-profit organization operated as a scientific partnership among the California Institute of Technology, the University of California, and the National Aeronautics and Space Administration. The Observatory was made possible by the generous financial support of the W. M. Keck Foundation. The authors wish to recognize and acknowledge the very significant cultural role and reverence that the summit of Maunakea has always had within the Native Hawaiian community. We are most fortunate to have the opportunity to conduct observations from this mountain.
T.C.B. acknowledges partial support for this work from grant PHY 14-30152; Physics Frontier Center/JINA Center for the Evolution of the Elements (JINA-CEE), and OISE-1927130: The International Research Network for Nuclear Astrophysics (IReNA), awarded by the US National Science Foundation. 

Portions of this text were written with the assistance of ChatGPT \citep{openai}, especially in regard to English usage and grammar.

\end{acknowledgements}

\begin{contribution}
J.~Hong led the scientific analysis and manuscript writing.  E.~Kirby advised on the scientific content and writing. 
T.~Tang provided guidance and feedback on the random forest analysis. M.~Chiba, Y.~Komiyama, and I.~Ogami provided the HSC data and suggestions for improving the manuscript. L.~Henderson and E.~Kirby provided the DEIMOS measurements. T.~Beers contributed suggestions for improving the manuscript.
\end{contribution}

\vfill\eject

\gdef\thefigure{\thesection.\arabic{figure}}    

\bibliography{main}{}

\begin{thebibliography}{}
\expandafter\ifx\csname natexlab\endcsname\relax\def\natexlab#1{#1}\fi
\providecommand{\url}[1]{\href{#1}{#1}}
\providecommand{\dodoi}[1]{doi:~\href{http://doi.org/#1}{\nolinkurl{#1}}}
\providecommand{\doeprint}[1]{\href{http://ascl.net/#1}{\nolinkurl{http://ascl.net/#1}}}
\providecommand{\doarXiv}[1]{\href{https://arxiv.org/abs/#1}{\nolinkurl{https://arxiv.org/abs/#1}}}

\bibitem[{{Barbosa} {et~al.}(2025){Barbosa}, {Chiti}, {Limberg}, {Pace}, {Cerny}, {Rossi}, {Carlin}, {Stringfellow}, {Placco}, {Atzerberg}, {Carballo-Bello}, {Chaturvedi}, {Choi}, {Crnojevic}, {Drlica-Wagner}, {Ji}, {Kallivayalil}, {Mart{\'\i}nez-V{\'a}zquez}, {Medina}, {Noel}, {Sand}, {Vivas}, {Bom}, {Ferguson}, {Mutlu-Pakdil}, {Navabi}, {Riley}, {Sakowska}, \& {Zenteno}}]{Barbosa2025}
{Barbosa}, F.~O., {Chiti}, A., {Limberg}, G., {et~al.} 2025, arXiv e-prints, arXiv:2504.03593, \dodoi{10.48550/arXiv.2504.03593}

\bibitem[{{Battaglia} {et~al.}(2006){Battaglia}, {Tolstoy}, {Helmi}, {Irwin}, {Letarte}, {Jablonka}, {Hill}, {Venn}, {Shetrone}, {Arimoto}, {Primas}, {Kaufer}, {Francois}, {Szeifert}, {Abel}, \& {Sadakane}}]{Battaglia2006}
{Battaglia}, G., {Tolstoy}, E., {Helmi}, A., {et~al.} 2006, \aap, 459, 423, \dodoi{10.1051/0004-6361:20065720}

\bibitem[{{Benitez} {et~al.}(2014){Benitez}, {Dupke}, {Moles}, {Sodre}, {Cenarro}, {Marin-Franch}, {Taylor}, {Cristobal}, {Fernandez-Soto}, {Mendes de Oliveira}, {Cepa-Nogue}, {Abramo}, {Alcaniz}, {Overzier}, {Hernandez-Monteagudo}, {Alfaro}, {Kanaan}, {Carvano}, {Reis}, {Martinez Gonzalez}, {Ascaso}, {Ballesteros}, {Xavier}, {Varela}, {Ederoclite}, {Vazquez Ramio}, {Broadhurst}, {Cypriano}, {Angulo}, {Diego}, {Zandivarez}, {Diaz}, {Melchior}, {Umetsu}, {Spinelli}, {Zitrin}, {Coe}, {Yepes}, {Vielva}, {Sahni}, {Marcos-Caballero}, {Kitaura}, {Maroto}, {Masip}, {Tsujikawa}, {Carneiro}, {Gonzalez Nuevo}, {Carvalho}, {Reboucas}, {Carvalho}, {Abdalla}, {Bernui}, {Pigozzo}, {Ferreira}, {Chandrachani Devi}, {Bengaly}, {Campista}, {Amorim}, {Asari}, {Bongiovanni}, {Bonoli}, {Bruzual}, {Cardiel}, {Cava}, {Cid Fernandes}, {Coelho}, {Cortesi}, {Delgado}, {Diaz Garcia}, {Espinosa}, {Galliano}, {Gonzalez-Serrano}, {Falcon-Barroso}, {Fritz}, {Fernandes}, {Gorgas}, {Hoyos}, {Jimenez-Teja}, {Lopez-Aguerri}, {Lopez-San Juan},
  {Mateus}, {Molino}, {Novais}, {OMill}, {Oteo}, {Perez-Gonzalez}, {Poggianti}, {Proctor}, {Ricciardelli}, {Sanchez-Blazquez}, {Storchi-Bergmann}, {Telles}, {Schoennell}, {Trujillo}, {Vazdekis}, {Viironen}, {Daflon}, {Aparicio-Villegas}, {Rocha}, {Ribeiro}, {Borges}, {Martins}, {Marcolino}, {Martinez-Delgado}, {Perez-Torres}, {Siffert}, {Calvao}, {Sako}, {Kessler}, {Alvarez-Candal}, {De Pra}, {Roig}, {Lazzaro}, {Gorosabel}, {Lopes de Oliveira}, {Lima-Neto}, {Irwin}, {Liu}, {Alvarez}, {Balmes}, {Chueca}, {Costa-Duarte}, {da Costa}, {Dantas}, {Diaz}, {Fabregat}, {Ferrari}, {Gavela}, {Gracia}, {Gruel}, {Gutierrez}, {Guzman}, {Hernandez-Fernandez}, {Herranz}, {Hurtado-Gil}, {Jablonsky}, {Laporte}, {Le Tiran}, {Licandro}, {Lima}, {Martin}, {Martinez}, {Montero}, {Penteado}, {Pereira}, {Peris}, {Quilis}, {Sanchez-Portal}, {Soja}, {Solano}, {Torra}, \& {Valdivielso}}]{Benitez2014}
{Benitez}, N., {Dupke}, R., {Moles}, M., {et~al.} 2014, arXiv e-prints, arXiv:1403.5237, \dodoi{10.48550/arXiv.1403.5237}

\bibitem[{{Bettinelli} {et~al.}(2019){Bettinelli}, {Hidalgo}, {Cassisi}, {Aparicio}, {Piotto}, {Valdes}, \& {Walker}}]{Bettinelli2019}
{Bettinelli}, M., {Hidalgo}, S.~L., {Cassisi}, S., {et~al.} 2019, \mnras, 487, 5862, \dodoi{10.1093/mnras/stz1679}

\bibitem[{{Biau} \& {Scornet}(2015)}]{Biau2015}
{Biau}, G., \& {Scornet}, E. 2015, arXiv e-prints, arXiv:1511.05741, \dodoi{10.48550/arXiv.1511.05741}

\bibitem[{{Breiman}(2001)}]{Breiman2001}
{Breiman}, L. 2001, Machine Learning, 45, 5, \dodoi{10.1023/A:1010933404324}

\bibitem[{{Breiman} {et~al.}(1984){Breiman}, {Friedman}, {Olshen}, \& {Stone}}]{Breiman1984}
{Breiman}, L., {Friedman}, J., {Olshen}, R., \& {Stone}, C.~J. 1984, Chapman and Hall/CRC

\bibitem[{{Cenarro} {et~al.}(2019){Cenarro}, {Moles}, {Crist{\'o}bal-Hornillos}, {Mar{\'\i}n-Franch}, {Ederoclite}, {Varela}, {L{\'o}pez-Sanjuan}, {Hern{\'a}ndez-Monteagudo}, {Angulo}, {V{\'a}zquez Rami{\'o}}, {Viironen}, {Bonoli}, {Orsi}, {Hurier}, {San Roman}, {Greisel}, {Vilella-Rojo}, {D{\'\i}az-Garc{\'\i}a}, {Logro{\~n}o-Garc{\'\i}a}, {Gurung-L{\'o}pez}, {Spinoso}, {Izquierdo-Villalba}, {Aguerri}, {Allende Prieto}, {Bonatto}, {Carvano}, {Chies-Santos}, {Daflon}, {Dupke}, {Falc{\'o}n-Barroso}, {Gon{\c{c}}alves}, {Jim{\'e}nez-Teja}, {Molino}, {Placco}, {Solano}, {Whitten}, {Abril}, {Ant{\'o}n}, {Bello}, {Bielsa de Toledo}, {Castillo-Ram{\'\i}rez}, {Chueca}, {Civera}, {D{\'\i}az-Mart{\'\i}n}, {Dom{\'\i}nguez-Mart{\'\i}nez}, {Garzar{\'a}n-Calderaro}, {Hern{\'a}ndez-Fuertes}, {Iglesias-Marzoa}, {I{\~n}iguez}, {Jim{\'e}nez Ruiz}, {Kruuse}, {Lamadrid}, {Lasso-Cabrera}, {L{\'o}pez-Alegre}, {L{\'o}pez-Sainz}, {Ma{\'\i}cas}, {Moreno-Signes}, {Muniesa}, {Rodr{\'\i}guez-Llano}, {Rueda-Teruel}, {Rueda-Teruel},
  {Soriano-Lagu{\'\i}a}, {Tilve}, {Valdivielso}, {Yanes-D{\'\i}az}, {Alcaniz}, {Mendes de Oliveira}, {Sodr{\'e}}, {Coelho}, {Lopes de Oliveira}, {Tamm}, {Xavier}, {Abramo}, {Akras}, {Alfaro}, {Alvarez-Candal}, {Ascaso}, {Beasley}, {Beers}, {Borges Fernandes}, {Bruzual}, {Buzzo}, {Carrasco}, {Cepa}, {Cortesi}, {Costa-Duarte}, {De Pr{\'a}}, {Favole}, {Galarza}, {Galbany}, {Garcia}, {Gonz{\'a}lez Delgado}, {Gonz{\'a}lez-Serrano}, {Guti{\'e}rrez-Soto}, {Hernandez-Jimenez}, {Kanaan}, {Kuncarayakti}, {Landim}, {Laur}, {Licandro}, {Lima Neto}, {Lyman}, {Ma{\'\i}z Apell{\'a}niz}, {Miralda-Escud{\'e}}, {Morate}, {Nogueira-Cavalcante}, {Novais}, {Oncins}, {Oteo}, {Overzier}, {Pereira}, {Rebassa-Mansergas}, {Reis}, {Roig}, {Sako}, {Salvador-Rusi{\~n}ol}, {Sampedro}, {S{\'a}nchez-Bl{\'a}zquez}, {Santos}, {Schmidtobreick}, {Siffert}, {Telles}, \& {Vilchez}}]{Cenarro2019}
{Cenarro}, A.~J., {Moles}, M., {Crist{\'o}bal-Hornillos}, D., {et~al.} 2019, \aap, 622, A176, \dodoi{10.1051/0004-6361/201833036}

\bibitem[{{Cohen} \& {Huang}(2009)}]{Cohen2009}
{Cohen}, J.~G., \& {Huang}, W. 2009, \apj, 701, 1053, \dodoi{10.1088/0004-637X/701/2/1053}

\bibitem[{{Cohen} \& {Huang}(2010)}]{Cohen2010}
---. 2010, \apj, 719, 931, \dodoi{10.1088/0004-637X/719/1/931}

\bibitem[{{Coleman} {et~al.}(2005){Coleman}, {Da Costa}, \& {Bland-Hawthorn}}]{Coleman2005}
{Coleman}, M.~G., {Da Costa}, G.~S., \& {Bland-Hawthorn}, J. 2005, \aj, 130, 1065, \dodoi{10.1086/432662}

\bibitem[{{de Boer} {et~al.}(2012{\natexlab{a}}){de Boer}, {Tolstoy}, {Hill}, {Saha}, {Olsen}, {Starkenburg}, {Lemasle}, {Irwin}, \& {Battaglia}}]{deBoer2012a}
{de Boer}, T.~J.~L., {Tolstoy}, E., {Hill}, V., {et~al.} 2012{\natexlab{a}}, \aap, 539, A103, \dodoi{10.1051/0004-6361/201118378}

\bibitem[{{de Boer} {et~al.}(2012{\natexlab{b}}){de Boer}, {Tolstoy}, {Hill}, {Saha}, {Olszewski}, {Mateo}, {Starkenburg}, {Battaglia}, \& {Walker}}]{deBoer2012b}
---. 2012{\natexlab{b}}, \aap, 544, A73, \dodoi{10.1051/0004-6361/201219547}

\bibitem[{{de los Reyes} {et~al.}(2022){de los Reyes}, {Kirby}, {Ji}, \& {Nu{\~n}ez}}]{delosReyes2022}
{de los Reyes}, M. A.~C., {Kirby}, E.~N., {Ji}, A.~P., \& {Nu{\~n}ez}, E.~H. 2022, \apj, 925, 66, \dodoi{10.3847/1538-4357/ac332b}

\bibitem[{{Edge} {et~al.}(2013){Edge}, {Sutherland}, {Kuijken}, {Driver}, {McMahon}, {Eales}, \& {Emerson}}]{Edge2013}
{Edge}, A., {Sutherland}, W., {Kuijken}, K., {et~al.} 2013, The Messenger, 154, 32

\bibitem[{{Fallows} \& {Sanders}(2022)}]{Fallows2022}
{Fallows}, C.~P., \& {Sanders}, J.~L. 2022, \mnras, 516, 5521, \dodoi{10.1093/mnras/stac2550}

\bibitem[{{Ferreira Lopes} {et~al.}(2025){Ferreira Lopes}, {Guti{\'e}rrez-Soto}, {Ferreira Alberice}, {Monsalves}, {Hazarika}, {Catelan}, {Placco}, {Limberg}, {Almeida-Fernandes}, {Perottoni}, {Smith Castelli}, {Akras}, {Alonso-Garc{\'\i}a}, {Cordeiro}, {Jaque Arancibia}, {Daflon}, {Dias}, {Gon{\c{c}}alves}, {Machado-Pereira}, {Lopes}, {Bom}, {Thom de Souza}, {de Is{\'\i}dio}, {Alvarez-Candal}, {De Rossi}, {Bonatto}, {Cubillos Palma}, {Borges Fernandes}, {Humire}, {Oliveira Schwarz}, {Schoenell}, {Kanaan}, \& {Mendes de Oliveira}}]{FerreiraLopes2025}
{Ferreira Lopes}, C.~E., {Guti{\'e}rrez-Soto}, L.~A., {Ferreira Alberice}, V.~S., {et~al.} 2025, \aap, 693, A306, \dodoi{10.1051/0004-6361/202451491}

\bibitem[{{Furusawa} {et~al.}(2018){Furusawa}, {Koike}, {Takata}, {Okura}, {Miyatake}, {Lupton}, {Bickerton}, {Price}, {Bosch}, {Yasuda}, {Mineo}, {Yamada}, {Miyazaki}, {Nakata}, {Koshida}, {Komiyama}, {Utsumi}, {Kawanomoto}, {Jeschke}, {Noumaru}, {Schubert}, {Iwata}, {Finet}, {Fujiyoshi}, {Tajitsu}, {Terai}, \& {Lee}}]{Furusawa2018}
{Furusawa}, H., {Koike}, M., {Takata}, T., {et~al.} 2018, \pasj, 70, S3, \dodoi{10.1093/pasj/psx079}

\bibitem[{{Han} {et~al.}(2020){Han}, {Kim}, {Yoon}, {Lee}, {Arimoto}, {Okamoto}, \& {Ree}}]{Han2020}
{Han}, S.-I., {Kim}, H.-S., {Yoon}, S.-J., {et~al.} 2020, \apjs, 247, 7, \dodoi{10.3847/1538-4365/ab6441}

\bibitem[{{Hasselquist} {et~al.}(2021){Hasselquist}, {Hayes}, {Lian}, {Weinberg}, {Zasowski}, {Horta}, {Beaton}, {Feuillet}, {Garro}, {Gallart}, {Smith}, {Holtzman}, {Minniti}, {Lacerna}, {Shetrone}, {J{\"o}nsson}, {Cioni}, {Fillingham}, {Cunha}, {O'Connell}, {Fern{\'a}ndez-Trincado}, {Mu{\~n}oz}, {Schiavon}, {Almeida}, {Anguiano}, {Beers}, {Bizyaev}, {Brownstein}, {Cohen}, {Frinchaboy}, {Garc{\'\i}a-Hern{\'a}ndez}, {Geisler}, {Lane}, {Majewski}, {Nidever}, {Nitschelm}, {Povick}, {Price-Whelan}, {Roman-Lopes}, {Rosado}, {Sobeck}, {Stringfellow}, {Valenzuela}, {Villanova}, \& {Vincenzo}}]{Hasselquist2021}
{Hasselquist}, S., {Hayes}, C.~R., {Lian}, J., {et~al.} 2021, \apj, 923, 172, \dodoi{10.3847/1538-4357/ac25f9}

\bibitem[{{Hastie} {et~al.}(2009){Hastie}, {Tibshirani}, \& {Friedman}}]{Hastie2009}
{Hastie}, T.~J., {Tibshirani}, R.~J., \& {Friedman}, J.~H. 2009, Springer

\bibitem[{{Henderson} {et~al.}(2025){Henderson}, {Kirby}, {de los Reyes}, {Gerasimov}, \& {Manwadkar}}]{Henderson2025}
{Henderson}, L.~E., {Kirby}, E.~N., {de los Reyes}, M. A.~C., {Gerasimov}, R., \& {Manwadkar}, V. 2025, \apj, 983, 117, \dodoi{10.3847/1538-4357/adbe7d}

\bibitem[{{Hendricks} {et~al.}(2014){Hendricks}, {Koch}, {Walker}, {Johnson}, {Pe{\~n}arrubia}, \& {Gilmore}}]{Hendricks2014}
{Hendricks}, B., {Koch}, A., {Walker}, M., {et~al.} 2014, \aap, 572, A82, \dodoi{10.1051/0004-6361/201424645}

\bibitem[{{Hill} {et~al.}(2019){Hill}, {Sk{\'u}lad{\'o}ttir}, {Tolstoy}, {Venn}, {Shetrone}, {Jablonka}, {Primas}, {Battaglia}, {de Boer}, {Fran{\c{c}}ois}, {Helmi}, {Kaufer}, {Letarte}, {Starkenburg}, \& {Spite}}]{Hill2019}
{Hill}, V., {Sk{\'u}lad{\'o}ttir}, {\'A}., {Tolstoy}, E., {et~al.} 2019, \aap, 626, A15, \dodoi{10.1051/0004-6361/201833950}

\bibitem[{{Hirai} {et~al.}(2024){Hirai}, {Kirby}, {Chiba}, {Hayashi}, {Anguiano}, {Saitoh}, {Ishigaki}, \& {Beers}}]{Hirai2024}
{Hirai}, Y., {Kirby}, E.~N., {Chiba}, M., {et~al.} 2024, \apj, 970, 105, \dodoi{10.3847/1538-4357/ad500c}

\bibitem[{{Huang} {et~al.}(2024){Huang}, {Beers}, {Xiao}, {Yuan}, {Lee}, {Gu}, {Hong}, {Liu}, {Fan}, {Coelho}, {Cruz}, {Galindo-Guil}, {Daflon}, {Jim{\'e}nez-Esteban}, {Cenarro}, {Crist{\'o}bal-Hornillos}, {Hern{\'a}ndez-Monteagudo}, {L{\'o}pez-Sanjuan}, {Mar{\'\i}n-Franch}, {Moles}, {Varela}, {V{\'a}zquez Rami{\'o}}, {Alcaniz}, {Dupke}, {Ederoclite}, {Sodr{\'e}}, \& {Angulo}}]{Huang2024}
{Huang}, Y., {Beers}, T.~C., {Xiao}, K., {et~al.} 2024, \apj, 974, 192, \dodoi{10.3847/1538-4357/ad6b94}

\bibitem[{{Jin} {et~al.}(2016){Jin}, {Irwin}, {Tolstoy}, {Lewis}, \& {Hartke}}]{Jin2016}
{Jin}, S., {Irwin}, M., {Tolstoy}, E., {Lewis}, J., \& {Hartke}, J. 2016, in Astronomical Society of the Pacific Conference Series, Vol. 507, Multi-Object Spectroscopy in the Next Decade: Big Questions, Large Surveys, and Wide Fields, ed. I.~{Skillen}, M.~{Balcells}, \& S.~{Trager}, 241

\bibitem[{{Kawanomoto} {et~al.}(2018){Kawanomoto}, {Uraguchi}, {Komiyama}, {Miyazaki}, {Furusawa}, {Finet}, {Hattori}, {Wang}, {Yasuda}, \& {Suzuki}}]{Kawanomoto2018}
{Kawanomoto}, S., {Uraguchi}, F., {Komiyama}, Y., {et~al.} 2018, \pasj, 70, 66, \dodoi{10.1093/pasj/psy056}

\bibitem[{{Kirby} {et~al.}(2013){Kirby}, {Cohen}, {Guhathakurta}, {Cheng}, {Bullock}, \& {Gallazzi}}]{Kirby2013}
{Kirby}, E.~N., {Cohen}, J.~G., {Guhathakurta}, P., {et~al.} 2013, \apj, 779, 102, \dodoi{10.1088/0004-637X/779/2/102}

\bibitem[{{Kirby} {et~al.}(2011{\natexlab{a}}){Kirby}, {Cohen}, {Smith}, {Majewski}, {Sohn}, \& {Guhathakurta}}]{Kirby2011b}
{Kirby}, E.~N., {Cohen}, J.~G., {Smith}, G.~H., {et~al.} 2011{\natexlab{a}}, \apj, 727, 79, \dodoi{10.1088/0004-637X/727/2/79}

\bibitem[{{Kirby} {et~al.}(2011{\natexlab{b}}){Kirby}, {Lanfranchi}, {Simon}, {Cohen}, \& {Guhathakurta}}]{Kirby2011a}
{Kirby}, E.~N., {Lanfranchi}, G.~A., {Simon}, J.~D., {Cohen}, J.~G., \& {Guhathakurta}, P. 2011{\natexlab{b}}, \apj, 727, 78, \dodoi{10.1088/0004-637X/727/2/78}

\bibitem[{{Kirby} {et~al.}(2010){Kirby}, {Guhathakurta}, {Simon}, {Geha}, {Rockosi}, {Sneden}, {Cohen}, {Sohn}, {Majewski}, \& {Siegel}}]{Kirby2010}
{Kirby}, E.~N., {Guhathakurta}, P., {Simon}, J.~D., {et~al.} 2010, \apjs, 191, 352, \dodoi{10.1088/0067-0049/191/2/352}

\bibitem[{{Komiyama} {et~al.}(2018{\natexlab{a}}){Komiyama}, {Obuchi}, {Nakaya}, {Kamata}, {Kawanomoto}, {Utsumi}, {Miyazaki}, {Uraguchi}, {Furusawa}, {Morokuma}, {Uchida}, {Miyatake}, {Mineo}, {Fujimori}, {Aihara}, {Karoji}, {Gunn}, \& {Wang}}]{Komiyama2018b}
{Komiyama}, Y., {Obuchi}, Y., {Nakaya}, H., {et~al.} 2018{\natexlab{a}}, \pasj, 70, S2, \dodoi{10.1093/pasj/psx069}

\bibitem[{{Komiyama} {et~al.}(2018{\natexlab{b}}){Komiyama}, {Chiba}, {Tanaka}, {Tanaka}, {Kirihara}, {Miki}, {Mori}, {Lupton}, {Guhathakurta}, {Kalirai}, {Gilbert}, {Kirby}, {Lee}, {Jang}, {Sharma}, \& {Hayashi}}]{Komiyama2018a}
{Komiyama}, Y., {Chiba}, M., {Tanaka}, M., {et~al.} 2018{\natexlab{b}}, \apj, 853, 29, \dodoi{10.3847/1538-4357/aaa129}

\bibitem[{{Kuijken} {et~al.}(2019){Kuijken}, {Heymans}, {Dvornik}, {Hildebrandt}, {de Jong}, {Wright}, {Erben}, {Bilicki}, {Giblin}, {Shan}, {Getman}, {Grado}, {Hoekstra}, {Miller}, {Napolitano}, {Paolilo}, {Radovich}, {Schneider}, {Sutherland}, {Tewes}, {Tortora}, {Valentijn}, \& {Verdoes Kleijn}}]{Kuijken2019}
{Kuijken}, K., {Heymans}, C., {Dvornik}, A., {et~al.} 2019, \aap, 625, A2, \dodoi{10.1051/0004-6361/201834918}

\bibitem[{{Lemasle} {et~al.}(2014){Lemasle}, {de Boer}, {Hill}, {Tolstoy}, {Irwin}, {Jablonka}, {Venn}, {Battaglia}, {Starkenburg}, {Shetrone}, {Letarte}, {Fran{\c{c}}ois}, {Helmi}, {Primas}, {Kaufer}, \& {Szeifert}}]{Lemasle2014}
{Lemasle}, B., {de Boer}, T.~J.~L., {Hill}, V., {et~al.} 2014, \aap, 572, A88, \dodoi{10.1051/0004-6361/201423919}

\bibitem[{{Letarte} {et~al.}(2010){Letarte}, {Hill}, {Tolstoy}, {Jablonka}, {Shetrone}, {Venn}, {Spite}, {Irwin}, {Battaglia}, {Helmi}, {Primas}, {Fran{\c{c}}ois}, {Kaufer}, {Szeifert}, {Arimoto}, \& {Sadakane}}]{Letarte2010}
{Letarte}, B., {Hill}, V., {Tolstoy}, E., {et~al.} 2010, \aap, 523, A17, \dodoi{10.1051/0004-6361/200913413}

\bibitem[{{Lianou} {et~al.}(2011){Lianou}, {Grebel}, \& {Koch}}]{Lianou2011}
{Lianou}, S., {Grebel}, E.~K., \& {Koch}, A. 2011, \aap, 531, A152, \dodoi{10.1051/0004-6361/201116998}

\bibitem[{Lu \& Hardin(2021)}]{lu2021unified}
Lu, B., \& Hardin, J. 2021, Journal of Machine Learning Research, 22, 1

\bibitem[{{Lucchesi} {et~al.}(2024){Lucchesi}, {Jablonka}, {Sk{\'u}lad{\'o}ttir}, {Lardo}, {Mashonkina}, {Primas}, {Venn}, {Hill}, \& {Minniti}}]{Lucchesi2024}
{Lucchesi}, R., {Jablonka}, P., {Sk{\'u}lad{\'o}ttir}, {\'A}., {et~al.} 2024, \aap, 686, A266, \dodoi{10.1051/0004-6361/202348093}

\bibitem[{{Majewski} {et~al.}(2000){Majewski}, {Ostheimer}, {Kunkel}, \& {Patterson}}]{Majewski2000}
{Majewski}, S.~R., {Ostheimer}, J.~C., {Kunkel}, W.~E., \& {Patterson}, R.~J. 2000, \aj, 120, 2550, \dodoi{10.1086/316836}

\bibitem[{{Maoz} {et~al.}(2010){Maoz}, {Sharon}, \& {Gal-Yam}}]{Maoz2010a}
{Maoz}, D., {Sharon}, K., \& {Gal-Yam}, A. 2010, \apj, 722, 1879, \dodoi{10.1088/0004-637X/722/2/1879}

\bibitem[{{Marcolini} {et~al.}(2006){Marcolini}, {D'Ercole}, {Brighenti}, \& {Recchi}}]{Marcolini2006}
{Marcolini}, A., {D'Ercole}, A., {Brighenti}, F., \& {Recchi}, S. 2006, \mnras, 371, 643, \dodoi{10.1111/j.1365-2966.2006.10671.x}

\bibitem[{{Mart{\'\i}nez-Delgado} {et~al.}(2001){Mart{\'\i}nez-Delgado}, {Alonso-Garc{\'\i}a}, {Aparicio}, \& {G{\'o}mez-Flechoso}}]{Martinez-Delgado2001}
{Mart{\'\i}nez-Delgado}, D., {Alonso-Garc{\'\i}a}, J., {Aparicio}, A., \& {G{\'o}mez-Flechoso}, M.~A. 2001, \apjl, 549, L63, \dodoi{10.1086/319150}

\bibitem[{{Mason} {et~al.}(2024){Mason}, {Crain}, {Schiavon}, {Weinberg}, {Pfeffer}, {Schaye}, {Schaller}, \& {Theuns}}]{Mason2024}
{Mason}, A.~C., {Crain}, R.~A., {Schiavon}, R.~P., {et~al.} 2024, \mnras, 533, 184, \dodoi{10.1093/mnras/stae1743}

\bibitem[{{McConnachie}(2012)}]{McConnachie2012}
{McConnachie}, A.~W. 2012, \aj, 144, 4, \dodoi{10.1088/0004-6256/144/1/4}

\bibitem[{{McConnachie} \& {Venn}(2020)}]{McConnachie2020}
{McConnachie}, A.~W., \& {Venn}, K.~A. 2020, \aj, 160, 124, \dodoi{10.3847/1538-3881/aba4ab}

\bibitem[{{McWilliam}(1997)}]{McWilliam1997}
{McWilliam}, A. 1997, \araa, 35, 503, \dodoi{10.1146/annurev.astro.35.1.503}

\bibitem[{{Mendes de Oliveira} {et~al.}(2019){Mendes de Oliveira}, {Ribeiro}, {Schoenell}, {Kanaan}, {Overzier}, {Molino}, {Sampedro}, {Coelho}, {Barbosa}, {Cortesi}, {Costa-Duarte}, {Herpich}, {Hernandez-Jimenez}, {Placco}, {Xavier}, {Abramo}, {Saito}, {Chies-Santos}, {Ederoclite}, {Lopes de Oliveira}, {Gon{\c{c}}alves}, {Akras}, {Almeida}, {Almeida-Fernandes}, {Beers}, {Bonatto}, {Bonoli}, {Cypriano}, {Vinicius-Lima}, {de Souza}, {Fabiano de Souza}, {Ferrari}, {Gon{\c{c}}alves}, {Gonzalez}, {Guti{\'e}rrez-Soto}, {Hartmann}, {Jaffe}, {Kerber}, {Lima-Dias}, {Lopes}, {Menendez-Delmestre}, {Nakazono}, {Novais}, {Ortega-Minakata}, {Pereira}, {Perottoni}, {Queiroz}, {Reis}, {Santos}, {Santos-Silva}, {Santucci}, {Barbosa}, {Siffert}, {Sodr{\'e}}, {Torres-Flores}, {Westera}, {Whitten}, {Alcaniz}, {Alonso-Garc{\'\i}a}, {Alencar}, {Alvarez-Candal}, {Amram}, {Azanha}, {Barb{\'a}}, {Bernardinelli}, {Borges Fernandes}, {Branco}, {Brito-Silva}, {Buzzo}, {Caffer}, {Campillay}, {Cano}, {Carvano}, {Castejon}, {Cid
  Fernandes}, {Dantas}, {Daflon}, {Damke}, {de la Reza}, {de Melo de Azevedo}, {De Paula}, {Diem}, {Donnerstein}, {Dors}, {Dupke}, {Eikenberry}, {Escudero}, {Faifer}, {Far{\'\i}as}, {Fernandes}, {Fernandes}, {Fontes}, {Galarza}, {Hirata}, {Katena}, {Gregorio-Hetem}, {Hern{\'a}ndez-Fern{\'a}ndez}, {Izzo}, {Jaque Arancibia}, {Jatenco-Pereira}, {Jim{\'e}nez-Teja}, {Kann}, {Krabbe}, {Labayru}, {Lazzaro}, {Lima Neto}, {Lopes}, {Magalh{\~a}es}, {Makler}, {de Menezes}, {Miralda-Escud{\'e}}, {Monteiro-Oliveira}, {Montero-Dorta}, {Mu{\~n}oz-Elgueta}, {Nemmen}, {Nilo Castell{\'o}n}, {Oliveira}, {Ort{\'\i}z}, {Pattaro}, {Pereira}, {Quint}, {Riguccini}, {Rocha Pinto}, {Rodrigues}, {Roig}, {Rossi}, {Saha}, {Santos}, {Schnorr M{\"u}ller}, {Sesto}, {Silva}, {Smith Castelli}, {Teixeira}, {Telles}, {Thom de Souza}, {Th{\"o}ne}, {Trevisan}, {de Ugarte Postigo}, {Urrutia-Viscarra}, {Veiga}, {Vika}, {Vitorelli}, {Werle}, {Werner}, \& {Zaritsky}}]{MendesdeOliveira2019}
{Mendes de Oliveira}, C., {Ribeiro}, T., {Schoenell}, W., {et~al.} 2019, \mnras, 489, 241, \dodoi{10.1093/mnras/stz1985}

\bibitem[{{Miyazaki} {et~al.}(2012){Miyazaki}, {Komiyama}, {Nakaya}, {Kamata}, {Doi}, {Hamana}, {Karoji}, {Furusawa}, {Kawanomoto}, {Morokuma}, {Ishizuka}, {Nariai}, {Tanaka}, {Uraguchi}, {Utsumi}, {Obuchi}, {Okura}, {Oguri}, {Takata}, {Tomono}, {Kurakami}, {Namikawa}, {Usuda}, {Yamanoi}, {Terai}, {Uekiyo}, {Yamada}, {Koike}, {Aihara}, {Fujimori}, {Mineo}, {Miyatake}, {Yasuda}, {Nishizawa}, {Saito}, {Tanaka}, {Uchida}, {Katayama}, {Wang}, {Chen}, {Lupton}, {Loomis}, {Bickerton}, {Price}, {Gunn}, {Suzuki}, {Miyazaki}, {Muramatsu}, {Yamamoto}, {Endo}, {Ezaki}, {Itoh}, {Miwa}, {Yokota}, {Matsuda}, {Ebinuma}, \& {Takeshi}}]{Miyazaki2012}
{Miyazaki}, S., {Komiyama}, Y., {Nakaya}, H., {et~al.} 2012, in Society of Photo-Optical Instrumentation Engineers (SPIE) Conference Series, Vol. 8446, Ground-based and Airborne Instrumentation for Astronomy IV, ed. I.~S. {McLean}, S.~K. {Ramsay}, \& H.~{Takami}, 84460Z, \dodoi{10.1117/12.926844}

\bibitem[{{Miyazaki} {et~al.}(2018){Miyazaki}, {Komiyama}, {Kawanomoto}, {Doi}, {Furusawa}, {Hamana}, {Hayashi}, {Ikeda}, {Kamata}, {Karoji}, {Koike}, {Kurakami}, {Miyama}, {Morokuma}, {Nakata}, {Namikawa}, {Nakaya}, {Nariai}, {Obuchi}, {Oishi}, {Okada}, {Okura}, {Tait}, {Takata}, {Tanaka}, {Tanaka}, {Terai}, {Tomono}, {Uraguchi}, {Usuda}, {Utsumi}, {Yamada}, {Yamanoi}, {Aihara}, {Fujimori}, {Mineo}, {Miyatake}, {Oguri}, {Uchida}, {Tanaka}, {Yasuda}, {Takada}, {Murayama}, {Nishizawa}, {Sugiyama}, {Chiba}, {Futamase}, {Wang}, {Chen}, {Ho}, {Liaw}, {Chiu}, {Ho}, {Lai}, {Lee}, {Jeng}, {Iwamura}, {Armstrong}, {Bickerton}, {Bosch}, {Gunn}, {Lupton}, {Loomis}, {Price}, {Smith}, {Strauss}, {Turner}, {Suzuki}, {Miyazaki}, {Muramatsu}, {Yamamoto}, {Endo}, {Ezaki}, {Ito}, {Kawaguchi}, {Sofuku}, {Taniike}, {Akutsu}, {Dojo}, {Kasumi}, {Matsuda}, {Imoto}, {Miwa}, {Suzuki}, {Takeshi}, \& {Yokota}}]{Miyazaki2018}
{Miyazaki}, S., {Komiyama}, Y., {Kawanomoto}, S., {et~al.} 2018, \pasj, 70, S1, \dodoi{10.1093/pasj/psx063}

\bibitem[{{Miyoshi} \& {Chiba}(2020)}]{Miyoshi2020}
{Miyoshi}, T., \& {Chiba}, M. 2020, \apj, 905, 109, \dodoi{10.3847/1538-4357/abc486}

\bibitem[{{Ogami} {et~al.}(2024){Ogami}, {Komiyama}, {Chiba}, {Tanaka}, {Guhathakurta}, {Kirby}, {Wyse}, {Filion}, {Kirihara}, {Ishigaki}, \& {Hayashi}}]{Ogami2024}
{Ogami}, I., {Komiyama}, Y., {Chiba}, M., {et~al.} 2024, \apj, 971, 107, \dodoi{10.3847/1538-4357/ad5445}

\bibitem[{{Ogami} {et~al.}(2025){Ogami}, {Tanaka}, {Komiyama}, {Chiba}, {Guhathakurta}, {Kirby}, {Wyse}, {Filion}, {Gilbert}, {Escala}, {Mori}, {Kirihara}, {Tanaka}, {Ishigaki}, {Hayashi}, {Lee}, {Sharma}, {Kalirai}, \& {Lupton}}]{Ogami2025}
{Ogami}, I., {Tanaka}, M., {Komiyama}, Y., {et~al.} 2025, \mnras, 536, 530, \dodoi{10.1093/mnras/stae2527}

\bibitem[{{{\"O}hman}(1934)}]{Ohman1934}
{{\"O}hman}, Y. 1934, \apj, 80, 171, \dodoi{10.1086/143595}

\bibitem[{{Okamoto} {et~al.}(2024){Okamoto}, {Ferguson}, {Arimoto}, {Ogami}, {{\v{Z}}emaitis}, {Chiba}, {Irwin}, {Jang}, {Koda}, {Komiyama}, {Lee}, {Lee}, {Rich}, {Tanaka}, \& {Tanaka}}]{Okamoto2024}
{Okamoto}, S., {Ferguson}, A. M.~N., {Arimoto}, N., {et~al.} 2024, \apjl, 967, L24, \dodoi{10.3847/2041-8213/ad4358}

\bibitem[{{OpenAI,}(2024)}]{openai}
{OpenAI,}. 2024, ChatGPT-4o (May 13 version) [Large Language Model], https://chat.openai.com/chat

\bibitem[{{Pace} {et~al.}(2022){Pace}, {Erkal}, \& {Li}}]{Pace2022}
{Pace}, A.~B., {Erkal}, D., \& {Li}, T.~S. 2022, \apj, 940, 136, \dodoi{10.3847/1538-4357/ac997b}

\bibitem[{{Pace} {et~al.}(2020){Pace}, {Kaplinghat}, {Kirby}, {Simon}, {Tollerud}, {Mu{\~n}oz}, {C{\^o}t{\'e}}, {Djorgovski}, \& {Geha}}]{Pace2020}
{Pace}, A.~B., {Kaplinghat}, M., {Kirby}, E., {et~al.} 2020, \mnras, 495, 3022, \dodoi{10.1093/mnras/staa1419}

\bibitem[{{Reichert} {et~al.}(2020){Reichert}, {Hansen}, {Hanke}, {Sk{\'u}lad{\'o}ttir}, {Arcones}, \& {Grebel}}]{Reichert2020}
{Reichert}, M., {Hansen}, C.~J., {Hanke}, M., {et~al.} 2020, \aap, 641, A127, \dodoi{10.1051/0004-6361/201936930}

\bibitem[{{Schlafly} {et~al.}(2019){Schlafly}, {Meisner}, \& {Green}}]{Schlafly2019}
{Schlafly}, E.~F., {Meisner}, A.~M., \& {Green}, G.~M. 2019, \apjs, 240, 30, \dodoi{10.3847/1538-4365/aafbea}

\bibitem[{Sch{\"o}lkopf {et~al.}(1998)Sch{\"o}lkopf, Smola, \& M{\"u}ller}]{scholkopf1998}
Sch{\"o}lkopf, B., Smola, A., \& M{\"u}ller, K.-R. 1998, Neural Computation, 10, 1299, \dodoi{10.1162/089976698300017467}

\bibitem[{{Sestito} {et~al.}(2023){Sestito}, {Zaremba}, {Venn}, {D'Aoust}, {Hayes}, {Jensen}, {Navarro}, {Jablonka}, {Fern{\'a}ndez-Alvar}, {Glover}, {McConnachie}, \& {Chen{\'e}}}]{Sestito2023}
{Sestito}, F., {Zaremba}, D., {Venn}, K.~A., {et~al.} 2023, \mnras, 525, 2875, \dodoi{10.1093/mnras/stad2427}

\bibitem[{{Shetrone} {et~al.}(2001){Shetrone}, {C{\^o}t{\'e}}, \& {Sargent}}]{Shetrone2001}
{Shetrone}, M.~D., {C{\^o}t{\'e}}, P., \& {Sargent}, W.~L.~W. 2001, \apj, 548, 592, \dodoi{10.1086/319022}

\bibitem[{{Skrutskie} {et~al.}(2006){Skrutskie}, {Cutri}, {Stiening}, {Weinberg}, {Schneider}, {Carpenter}, {Beichman}, {Capps}, {Chester}, {Elias}, {Huchra}, {Liebert}, {Lonsdale}, {Monet}, {Price}, {Seitzer}, {Jarrett}, {Kirkpatrick}, {Gizis}, {Howard}, {Evans}, {Fowler}, {Fullmer}, {Hurt}, {Light}, {Kopan}, {Marsh}, {McCallon}, {Tam}, {Van Dyk}, \& {Wheelock}}]{Skrutskie2006}
{Skrutskie}, M.~F., {Cutri}, R.~M., {Stiening}, R., {et~al.} 2006, \aj, 131, 1163, \dodoi{10.1086/498708}

\bibitem[{{Sk{\'u}lad{\'o}ttir} {et~al.}(2024){Sk{\'u}lad{\'o}ttir}, {Vanni}, {Salvadori}, \& {Lucchesi}}]{Skuladottir2024}
{Sk{\'u}lad{\'o}ttir}, {\'A}., {Vanni}, I., {Salvadori}, S., \& {Lucchesi}, R. 2024, \aap, 681, A44, \dodoi{10.1051/0004-6361/202346231}

\bibitem[{{Sk{\'u}lad{\'o}ttir} {et~al.}(2021){Sk{\'u}lad{\'o}ttir}, {Salvadori}, {Amarsi}, {Tolstoy}, {Irwin}, {Hill}, {Jablonka}, {Battaglia}, {Starkenburg}, {Massari}, {Helmi}, \& {Posti}}]{Skuladottir2021}
{Sk{\'u}lad{\'o}ttir}, {\'A}., {Salvadori}, S., {Amarsi}, A.~M., {et~al.} 2021, \apjl, 915, L30, \dodoi{10.3847/2041-8213/ac0dc2}

\bibitem[{{Starkenburg} {et~al.}(2010){Starkenburg}, {Hill}, {Tolstoy}, {Gonz{\'a}lez Hern{\'a}ndez}, {Irwin}, {Helmi}, {Battaglia}, {Jablonka}, {Tafelmeyer}, {Shetrone}, {Venn}, \& {de Boer}}]{Starkenburg2010}
{Starkenburg}, E., {Hill}, V., {Tolstoy}, E., {et~al.} 2010, \aap, 513, A34, \dodoi{10.1051/0004-6361/200913759}

\bibitem[{{Sun} {et~al.}(2025){Sun}, {Chen}, \& {Liu}}]{Sun2025}
{Sun}, B.-K., {Chen}, B.-Q., \& {Liu}, X.-W. 2025, Research in Astronomy and Astrophysics, 25, 035001, \dodoi{10.1088/1674-4527/adb55b}

\bibitem[{{Takada} {et~al.}(2014){Takada}, {Ellis}, {Chiba}, {Greene}, {Aihara}, {Arimoto}, {Bundy}, {Cohen}, {Dor{\'e}}, {Graves}, {Gunn}, {Heckman}, {Hirata}, {Ho}, {Kneib}, {Le F{\`e}vre}, {Lin}, {More}, {Murayama}, {Nagao}, {Ouchi}, {Seiffert}, {Silverman}, {Sodr{\'e}}, {Spergel}, {Strauss}, {Sugai}, {Suto}, {Takami}, \& {Wyse}}]{Takada2014}
{Takada}, M., {Ellis}, R.~S., {Chiba}, M., {et~al.} 2014, \pasj, 66, R1, \dodoi{10.1093/pasj/pst019}

\bibitem[{{Tang} {et~al.}(2023){Tang}, {Zhang}, {Yan}, {Zhang}, {Carigi}, \& {Fern{\'a}ndez-Trincado}}]{Tang2023}
{Tang}, B., {Zhang}, J., {Yan}, Z., {et~al.} 2023, \aap, 669, A125, \dodoi{10.1051/0004-6361/202244052}

\bibitem[{{Thackeray}(1939)}]{Thackeray1939}
{Thackeray}, A.~D. 1939, \mnras, 99, 492, \dodoi{10.1093/mnras/99.6.492}

\bibitem[{{Tolstoy} {et~al.}(2009){Tolstoy}, {Hill}, \& {Tosi}}]{Tolstoy2009}
{Tolstoy}, E., {Hill}, V., \& {Tosi}, M. 2009, \araa, 47, 371, \dodoi{10.1146/annurev-astro-082708-101650}

\bibitem[{{Tolstoy} {et~al.}(2023){Tolstoy}, {Sk{\'u}lad{\'o}ttir}, {Battaglia}, {Brown}, {Massari}, {Irwin}, {Starkenburg}, {Salvadori}, {Hill}, {Jablonka}, {Salaris}, {van Essen}, {Olsthoorn}, {Helmi}, \& {Pritchard}}]{Tolstoy2023}
{Tolstoy}, E., {Sk{\'u}lad{\'o}ttir}, {\'A}., {Battaglia}, G., {et~al.} 2023, \aap, 675, A49, \dodoi{10.1051/0004-6361/202245717}

\bibitem[{{Ural} {et~al.}(2015){Ural}, {Cescutti}, {Koch}, {Kleyna}, {Feltzing}, \& {Wilkinson}}]{Ural2015}
{Ural}, U., {Cescutti}, G., {Koch}, A., {et~al.} 2015, \mnras, 449, 761, \dodoi{10.1093/mnras/stv294}

\bibitem[{{Venn}(2014)}]{Venn2014}
{Venn}, K.~A. 2014, \memsai, 85, 519

\bibitem[{{Walker} {et~al.}(2009){Walker}, {Mateo}, \& {Olszewski}}]{Walker2009}
{Walker}, M.~G., {Mateo}, M., \& {Olszewski}, E.~W. 2009, \aj, 137, 3100, \dodoi{10.1088/0004-6256/137/2/3100}

\bibitem[{{Walker} {et~al.}(2007){Walker}, {Mateo}, {Olszewski}, {Gnedin}, {Wang}, {Sen}, \& {Woodroofe}}]{Walker2007}
{Walker}, M.~G., {Mateo}, M., {Olszewski}, E.~W., {et~al.} 2007, \apjl, 667, L53, \dodoi{10.1086/521998}

\bibitem[{{Weisz} {et~al.}(2014){Weisz}, {Dolphin}, {Skillman}, {Holtzman}, {Gilbert}, {Dalcanton}, \& {Williams}}]{Weisz2014}
{Weisz}, D.~R., {Dolphin}, A.~E., {Skillman}, E.~D., {et~al.} 2014, \apj, 789, 147, \dodoi{10.1088/0004-637X/789/2/147}

\bibitem[{{Wright} {et~al.}(2010){Wright}, {Eisenhardt}, {Mainzer}, {Ressler}, {Cutri}, {Jarrett}, {Kirkpatrick}, {Padgett}, {McMillan}, {Skrutskie}, {Stanford}, {Cohen}, {Walker}, {Mather}, {Leisawitz}, {Gautier}, {McLean}, {Benford}, {Lonsdale}, {Blain}, {Mendez}, {Irace}, {Duval}, {Liu}, {Royer}, {Heinrichsen}, {Howard}, {Shannon}, {Kendall}, {Walsh}, {Larsen}, {Cardon}, {Schick}, {Schwalm}, {Abid}, {Fabinsky}, {Naes}, \& {Tsai}}]{Wright2010}
{Wright}, E.~L., {Eisenhardt}, P. R.~M., {Mainzer}, A.~K., {et~al.} 2010, \aj, 140, 1868, \dodoi{10.1088/0004-6256/140/6/1868}

\bibitem[{{Yang} {et~al.}(2022){Yang}, {Yuan}, {Xiang}, {Duan}, {Huang}, {Liu}, {Beers}, {Galarza}, {Daflon}, {Fern{\'a}ndez-Ontiveros}, {Cenarro}, {Crist{\'o}bal-Hornillos}, {Hern{\'a}ndez-Monteagudo}, {L{\'o}pez-Sanjuan}, {Mar{\'\i}n-Franch}, {Moles}, {Varela}, {V{\'a}zquez Rami{\'o}}, {Alcaniz}, {Dupke}, {Ederoclite}, {Sodr{\'e}}, \& {Angulo}}]{Yang2022}
{Yang}, L., {Yuan}, H., {Xiang}, M., {et~al.} 2022, \aap, 659, A181, \dodoi{10.1051/0004-6361/202142724}

\end{thebibliography}
\bibliographystyle{aasjournal}

\end{document}